\documentclass[12pt]{article}
\usepackage{amsmath}
\usepackage{graphicx,psfrag,epsf}
\usepackage{enumerate}
\usepackage{natbib}
\usepackage{subfigure}
\usepackage{booktabs} 

\newcommand{\corr}     {\mbox{\tt corr}}

\newcommand{\cbold}    {\mbox{\bf c}}

\newcommand{\gbold}    {\mbox{\bf g}}

\newcommand{\xbold}    {\mbox{\bf x}}
\newcommand{\ybold}    {\mbox{\bf y}}

\newcommand{\Ibold}    {\mbox{\bf I}}

\newcommand{\Rbold}    {\mbox{\bf R}}

\newcommand{\betabold}    {\mbox{\boldmath${\beta}$}}
\newcommand{\epsilonbold} {\mbox{\boldmath${\epsilon}$}}

\newcommand{\Phibold}     {\mbox{\boldmath${\Phi}$}}

\newcommand{\thetabold}   {\mbox{\boldmath${\theta}$}}

\usepackage{amsfonts}
\usepackage{amssymb}
\usepackage{algorithm}
\usepackage{hyperref}
\usepackage[noend]{algpseudocode}
\usepackage{atbegshi}
\usepackage{color}
\usepackage{textcomp}

\newcommand{\blind}{0}

\newcounter{xxx}
\begin{document}

\def\spacingset#1{\renewcommand{\baselinestretch}%
{#1}\small\normalsize} \spacingset{1}


\if0\blind
{
  \title{\bf Adaptive semiparametric Bayesian differential equations via sequential Monte Carlo}
  \author{Shijia Wang
   \hspace{.2cm}\\
   School of Statistics and Data Science, LPMC \& KLMDASR, Nankai University, China\\
      Shufei Ge
   \hspace{.2cm}\\
   Institute of Mathematical Sciences, ShanghaiTech University, China\\  
   Renny Doig \\
   Department of Statistics and Actuarial Science, Simon Fraser University, BC, Canada\\  
   Liangliang Wang \\
   Department of Statistics and Actuarial Science, Simon Fraser University, BC, Canada}
  \maketitle
} \fi

\if1\blind
{
  \bigskip
  \bigskip
  \bigskip
  \begin{center}
    {\LARGE\bf Title}
\end{center}
  \medskip
} \fi

\bigskip
\begin{abstract}

Nonlinear differential equations (DEs) are used in a wide range of scientific problems to model complex dynamic systems. The differential equations often contain unknown parameters that are of scientific interest, which have to be estimated from noisy measurements of the dynamic system. Generally, there is no closed-form solution for nonlinear DEs, and the likelihood surface for the parameter of interest is multi-modal and very sensitive to different parameter values.  We propose a Bayesian framework for nonlinear {\color{black} DE systems}. A flexible nonparametric function is used to represent the dynamic process such that expensive numerical solvers can be avoided.  A sequential Monte Carlo {\color{black} algorithm} in the annealing framework is proposed to conduct Bayesian inference for parameters in DEs. In our numerical experiments, we use examples of ordinary differential equations and delay differential equations to demonstrate the effectiveness of the proposed algorithm. We developed an R package that is available at \url{https://github.com/shijiaw/smcDE}.

\end{abstract}

\noindent%

\spacingset{1.45}
\section{Introduction}
\label{sec:intro}

Nonlinear differential equations (\emph{e.g.}\ nonlinear ordinary or delay differential equations) are commonly used in modelling dynamic systems in ecology, physics, and engineering.  Delay differential equations (DDEs) are described by equations $d\xbold(t)/dt = \gbold(\xbold(t), \xbold(t-\tau)|\thetabold)$, where $\thetabold$ is the vector of unknown parameters and $\tau$ is the time delay parameter. These are continuous-time models for interactions between variables $\xbold(t)$ and a time delay $\tau$. Ordinary differential equations (ODEs) are often presented by 
$d\xbold(t)/dt = \gbold(\xbold(t)|\thetabold)$, which can be regarded as a special case of DDEs with $\tau = 0$. 
The form of $\gbold(\cdot)$ is generally proposed by specialists with scientific intuition. For example, ecologists proposed the simple Lotka-Volterra model \citep{rosenzweig1963graphical} to understand and predict the populations of predators and prey in an ecosystem. Given a concrete form of the function $\gbold(\xbold(t), \xbold(t-\tau)|\thetabold)$, the parameters $\thetabold$ and $\tau$ are unknown and need to be estimated using observations at some data points.  Differential equations (DEs) are often  observed with measurement error.  We assume  that the observed $\ybold(t)$  is linked to $\xbold(t)$ though an additive error model such that $\ybold(t) = \xbold(t) + \epsilonbold$, where $\epsilonbold$ is measurement error.   The estimation of parameters in DEs is of great interest and usually requires us to solve the DEs  $d\xbold(t)/dt = \gbold(\xbold(t), \xbold(t-\tau)|\thetabold)$.

Many DE systems do not admit an analytic solution.  One alternative approach is to solve the DEs numerically \citep{butcher2016numerical}, for example by using the Euler method \citep{jain1979numerical, bulirsch1966numerical}, the Exponential integrators \citep{hochbruck1998exponential, hochbruck2010exponential}, or  the Runge-Kutta method \citep{jameson1981numerical, ascher1997implicit}. However, numerical DE solvers are computationally expensive, especially for DDEs. Various methods have been proposed to solve DEs more efficiently in recent decades. The idea of using smoothing splines to fit dynamic data was first proposed by \cite{varah1982spline}. \cite{ramsay2007applied}, \cite{poyton2006parameter},  \cite{chen2008efficient} extended  the idea of smoothing to a two-stage approach. In the first stage, spline coefficients are optimized by minimizing the sum of the squared distances between the data and the spline functions at the observation times. In the second stage, using the estimated spline coefficients, DE parameters are optimized by minimizing the residuals of DE models. The two-stage approach may lead to inconsistent estimates \citep{ramsay2007parameter}. 
\cite{ramsay2007parameter} proposed 
a generalized smoothing approach, called ``parameter cascading'',  based on data smoothing methods and a generalization of profiled estimation. In the proposed approach, the spline coefficients are treated as nuisance parameters. Their method iterates between optimizing the objective function with respect to the spline coefficients given current DE parameter estimates, and optimizing the objective function with respect to the DE parameters given the estimated spline coefficients. The iteration is repeated until convergence is achieved. 
The parameter estimates are consistent and asymptotically normally distributed under mild conditions \citep{qi2010asymptotic, pang2017asymptotically}.  
There are several variations of the parameter cascading approach. \cite{cao2011robust} proposed a robust algorithm to estimate parameters using measurements with outliers based on smoothing splines. \cite{cao2012penalized} proposed a method to estimate time-varying parameters in ODEs, in which the ODE parameters are also modelled by smoothing splines.
 \cite{wang2012estimating} developed a semiparametric
method  with smoothing spline to estimate DDE parameters.

Using smoothing splines to model DEs is computationally  efficient since we do not need to numerically solve DEs. 
Most methods based on data smoothing to estimate parameters of DEs are derived from a frequentist perspective. Bayesian methods are of interest since they quantify the uncertainty of parameters. \cite{campbell2012smooth}  proposed a 
smooth functional tempering algorithm to conduct posterior inference for ODE parameters. This idea originates from parallel tempering and model-based smoothing. 
\cite{zhang2017bayesian}  proposed a high-dimensional linear  ODE model to accommodate the  directional interaction between areas of the brain. Parallelized schemes for  Markov chain Monte Carlo have been proposed to estimate this model. \cite{bhaumik2015bayesian} investigated a two-stage procedure to estimate parameters by minimizing the penalized ODEs.

There are several lines of work involved in estimating DE parameters from a Bayesian  perspective based on numerical DE solvers.  
\cite{dass2017laplace} proposed a two-step approach to approximate posterior distributions of parameters of interest. 
They first applied a numerical algorithm to solve ODEs, then integrated out nuisance parameters using Laplace approximations.
 \cite{bhaumik2017efficient} proposed a modification of \cite{bhaumik2015bayesian} by directly considering the distance between the function in the nonparametric
model and that obtained from a four-stage Runge-Kutta (RK4) method.
\cite{calderhead2009accelerating} presented a novel Bayesian sampler to infer parameters in nonlinear delay differential equations{\color{black};} the derivatives and time delay parameters were estimated via Gaussian processes. To make the DE estimation more consistent, 
\cite{dondelinger2013ode} proposed  an adaptive gradient matching approach  to jointly infer the hyperparameters of
a  Gaussian process as well as ODE parameters. 
\cite{barber2014gaussian}  simplified previous approaches by proposing a more natural generative model via 
Gaussian process. The proposed approach directly
links state derivative information with system observations.

Standard sequential Monte Carlo (SMC) methods \citep{doucet2001introduction, doucet2000sequential, liu1998sequential} are popular approaches for estimating dynamic models (e.g.\ state space models).  
SMC methods combine importance sampling and resampling algorithms. Under mild conditions, consistency properties and asymptotic normality hold \citep{chopin2004central}.  
\cite{del2006sequential} proposed a general SMC framework, to sample sequentially from a sequence of intermediate probability distributions that are defined on a common space. 
This general framework has promoted popularity of  SMC methods in  areas besides state space models. For example,  \cite{wang2018annealed} proposed an annealed SMC algorithm for phylogenetics  by designing an  
artificial sequence of intermediate distributions. 
Several SMC methods have been proposed to estimate parameters in ODE models. 
\cite{zhou2016toward} presented an adaptive SMC sampling strategy to estimate parameters and conduct model selection.  They used a simple example of ODEs to demonstrate the performance of model selection using their  algorithm. 
\cite{lee2018inference} introduced additive Gaussian errors into the ODE trajectory provided by numerical solvers, and they proposed a particle filter to infer ODE parameters. In addition, Gaussian processes have been used to avoid numerical integration. These works are based on numerically solving ODE models.

In this article, we propose a 
semiparametric Bayesian model for nonlinear DEs and design an  annealed SMC algorithm to conduct inference efficiently for parameters. The  DE trajectories are represented using a linear combination of basis functions. Consequently, our method avoids expensive numerical solvers, especially those for DDEs. It instead needs to estimate the basis coefficients together with other parameters in the DEs. In other words,   the parameters of interest include the DE parameters, basis coefficients of smoothing spline functions, and parameters in the observation model.  In addition, a tuning parameter is used to balance the fit to data and fidelity to the DEs. We  estimate the tuning parameter  using the Bayesian approach to avoid tuning it through expensive cross-validation. Inspired by the reference distribution of \cite{fan2011choosing} in the context of model selection,  we design an artificial sequence of intermediate distributions that starts from a reference distribution, which is easier to sample from, and gradually approaches the target distribution through a sequence of annealing parameters. 
The proposed annealed SMC can effectively sample parameters with multiple isolated posterior modes and basis function coefficients of  high dimensionality.  It adopts the adaptive scheme in \cite{zhou2016toward} and \cite{wang2018annealed}  to choose  the sequence of annealing parameters that determines  the intermediate target distributions of SMC.  Our numerical experiments demonstrate the effectiveness of our algorithm in estimating parameters and DE trajectories for  both ODEs and DDEs.

The rest of article is organized as follows. In \emph{Section}  \ref{sec:2}, we construct a fully Bayesian framework for nonlinear DEs.  In \emph{Section}  \ref{sec:3}, we introduce our new algorithm for Bayesian inference for nonlinear DEs. In \emph{Section}  \ref{sec:real} and  \emph{Section}  \ref{sec:simulation}, we use a real data analysis and  numerical experiments to show the effectiveness of our method. We conclude in \emph{Section}  \ref{sec:conc}.

\section{Hierarchical Bayesian differential equations}
\label{sec:2}
In this section, we introduce a hierarchical Bayesian structure for DE models.  In \emph{Section} \ref{sec:delikelihood}, we introduce the likelihood function for DEs. In \emph{Section} \ref{sec: fullBayes}, we construct a 
 Bayesian model for the DE model. 
In \emph{Section} \ref{sec: lambda}, we introduce selection of the tuning parameter $\lambda$. {\color{black} In \emph{Section} \ref{sec:post2}, we introduce the posterior distribution of  the DE model.}

\subsection{DE models}
\label{sec:delikelihood}
 We use $\xbold(t) = (x_{1}(t), \ldots, x_{I}(t))'$ to denote the DE variables  (\emph{i.e.}\ the solution of a DE system), where 
$x_{i}(t)$ denotes the $i$-th DE variable and $I$ denotes the total number of DE variables. Each DE variable $x_{i}(t), i=1, \ldots, I$, is a dynamic process modelled with one differential equation
\begin{eqnarray}
\frac{dx_{i}(t)}{dt} &=& g_{i}(\xbold(t), \xbold(t-\tau)|\thetabold),\nonumber\\
x_{i}(0) &=& x_{i0},
\end{eqnarray}
where {\color{black} $t\in [t_1, t_{max}]$, $t_1=0$ unless it is specified otherwise},  $\thetabold$ denotes the vector of unknown parameters in the DE model, $\tau$ is the delay parameter in DDE model ($\tau = 0$ in ODE model), and $x_{i}(0)$ is the initial condition for the $i$-th DE variable, which is also unknown and needs to be estimated. 
Delay differential equations (DDEs) are time-delayed systems. The time delay $\tau$ in DDEs considers the dependence of the present state of the DE variable based on its past state. 
 We refer readers to \emph{Section} \ref{sec:real} for a more detailed description of DDE models. 

We do not observe the DEs directly, instead we observe them with measurement error. \textcolor{black}{In addition, we often only observe a subset of the $I$ DE variables,  $\mathcal{I}_0 \subseteq \{1, \ldots, I\}$}. We let $\ybold_{i} = (y_{i1}, \ldots, y_{i{\color{black} J_i}})'$ denote the observations for the $i$-th DE trajectory. The $j$-th observation of $\ybold_{i}$ is assumed to be normally distributed with mean $x_{i}(t_{ij}|\thetabold, \tau, x_{i0})$ and variance $\sigma_{i}^{2}$, 
\[
y_{ij} \sim N(x_{i}(t_{ij}|\thetabold, \tau, x_{i0}), \sigma_{i}^{2}), j=1, \ldots, {\color{black} J_i},
\]
where $x_{i}(t_{ij}|\thetabold, \tau, x_{i0})$ denotes the DE solution given $\thetabold$, $\tau$, and initial condition $x_{i0}$; $t_{ij}$ denotes the time we observe the $j$-th observation of $\ybold_{i}$.

\begin{figure}[ht]
\centering
\subfigure[]{
  \includegraphics[scale=0.4]{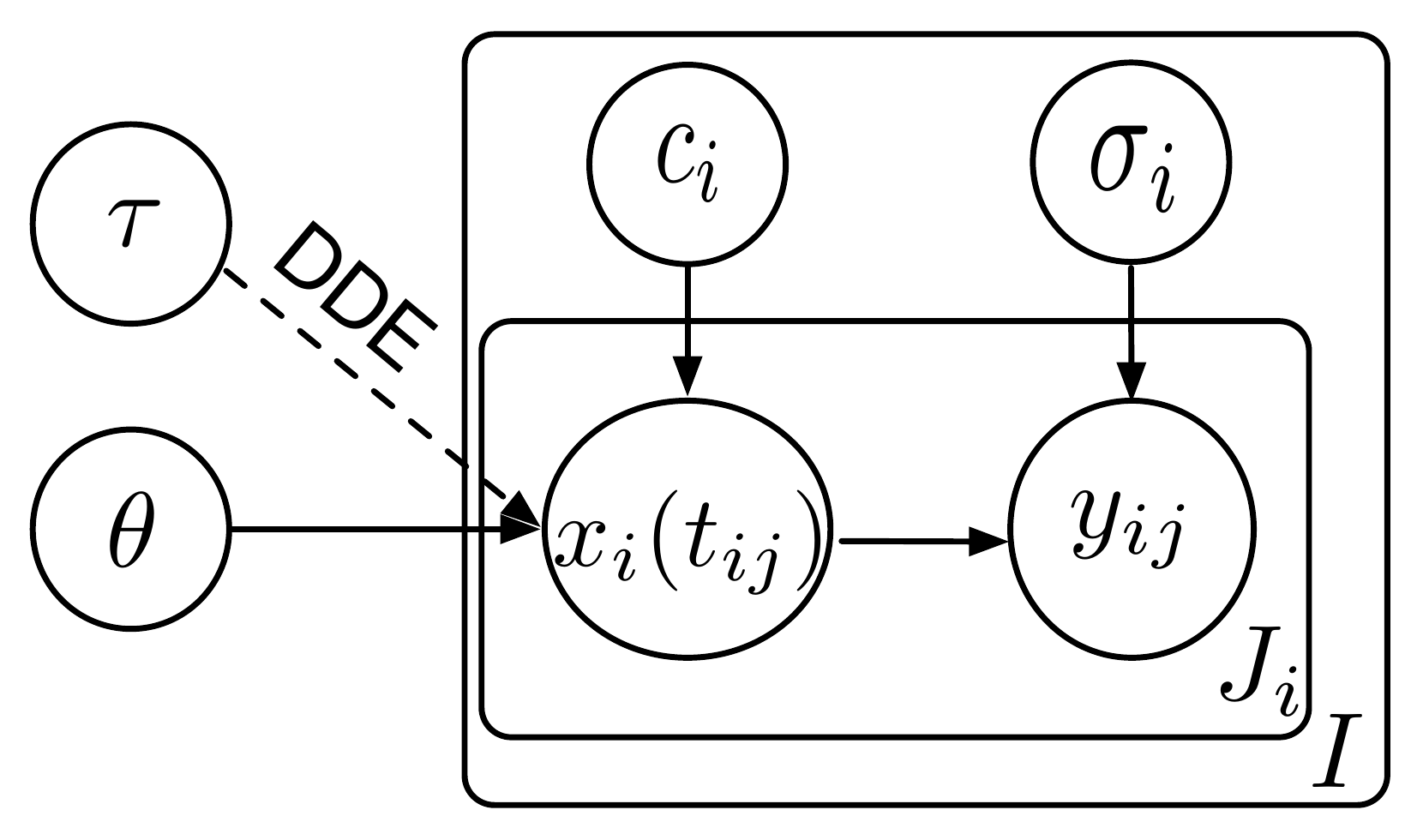}
    \label{fig:subfig1}
}
\subfigure[]{
\includegraphics[scale=0.35]{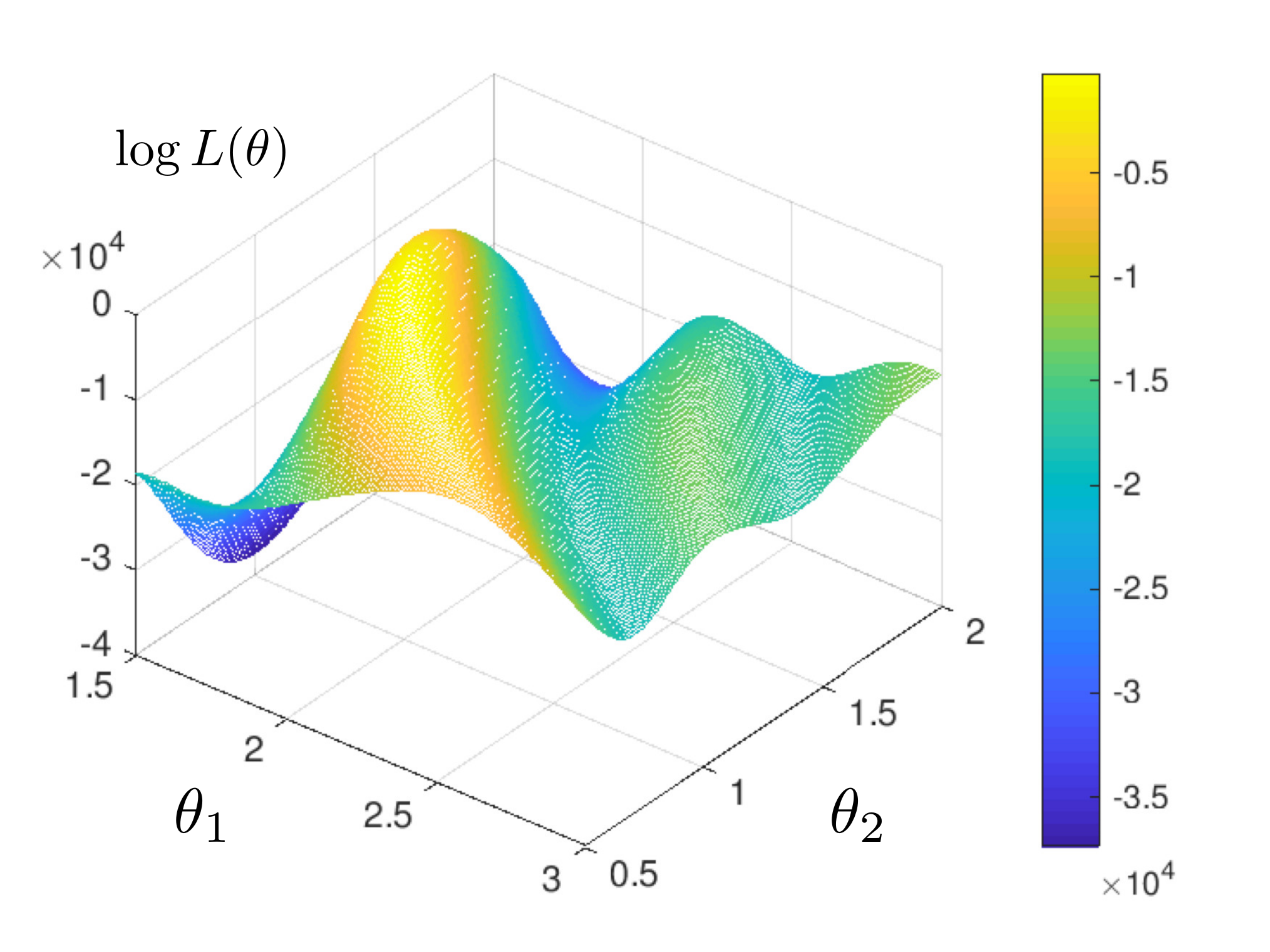}
    \label{fig:subfig2}
}
\caption[Optional caption for list of figures]{\subref{fig:subfig1} Graphical representation of DEs; \subref{fig:subfig2} Log-likelihood surface for a DE model.}
\label{fig:graphandLL}
\end{figure}

The joint likelihood function of $\thetabold$, $\tau$, $\xbold(0)$, and $\sigma_{i}^{2}$ admits the following form
\begin{equation}
\label{eq:1}
L(\thetabold, \tau, \xbold(0), \sigma_{i}^{2}) = \prod_{\textcolor{black}{i\in \mathcal{I}_0 }} \prod_{j=1}^{{\color{black} J_i}}(\sigma_{i}^{2})^{-1/2}\exp\bigg\{ -\frac{(y_{ij} - x_{i}(t_{ij}|\thetabold, \tau, \xbold(0)))^{2}}{2\sigma_{i}^{2}} \bigg\}.
\end{equation}
We use a figure (see Figure \ref{fig:graphandLL} (b)) to show an example of the log-likelihood surface over the DE parameters $\thetabold$, and for the setup of this model we refer to {\it Section \ref{sec:sim1}}.
The log-likelihood surface for $\thetabold$ has multiple isolated modes, and it is very sensitive to different parameter values. 

\subsection{A Bayesian structure for DE model}
\label{sec: fullBayes}

Numerically solving DEs can be computationally extremely intensive, especially for DDE models. 
We propose to solve differential equations by penalized smoothing. More specifically, we represent the $i$-th DE function $x_{i}(t)$ as a linear combination of $L_{i}$ B-spline basis functions {\color{black} $\Phibold_i(t) = (\phi_{i1}(t), \phi_{i2}(t), \ldots, \phi_{iL_{i}}(t))'$},
\begin{eqnarray*}
x_{i}(t) = \Phibold_i(t)' \cbold_{i},
\end{eqnarray*}
where $\cbold_{i}$ denotes the vector of basis coefficients.  See Figure  $1$ in {\it Supplementary \textcolor{black}{Material}} for an example of cubic B-spline functions \citep{ramsay2004functional, de1972calculating}.
Here we do not distinguish  the true $x_{i}(t)$ from its approximation using splines, which is a common practice in nonparametric smoothing \citep{berry2002bayesian,ramsay2004functional, wood2017generalized}. {\color{black}  The reason is that the number of basis functions is chosen to be sufficiently large such that we expect the basis function approximation can avoid bias from model over-simplification. 
The error in approximating $x_i(t)$ by its B-spline approximation typically is negligible compared to the estimation error, so we can assume that the two are equivalent (See Page 161 of \cite{berry2002bayesian}).} The initial condition for the $i$-th DE function is $x_{i}(0) = \Phibold_i(0)'\cbold_{i}$. One advantage of using smoothing spline functions to model DE trajectories is that we can avoid  explicitly estimating  the initial condition $\xbold(0)$; instead, it is estimated using {\color{black} $\hat{x}_{i}(0) = \Phibold_i(0)' \hat{\cbold}_{i}$}, where $\hat{\cbold}_{i}$ is the vector of estimated basis coefficients. Figure \ref{fig:graphandLL} (a) represents the graphical structure for the proposed DE model. The unknown parameters in our DE model include spline coefficients $\cbold_{i}$, the delay time parameter $\tau$ (which is known in an ODE with $\tau = 0$), the DE parameter $\thetabold$, and variance parameter $\sigma_{i}^{2}$.

With the basis function representation, finding the DE solution becomes a problem of estimating the basis function coefficients, {\color{black} $\cbold = (\cbold'_1, \cbold'_2, \ldots, \cbold'_I)'$}. In the Bayesian framework, we specify a prior distribution for $\cbold$ conditional on the DE parameters $\thetabold$, $\tau$, and a smoothing parameter $\lambda$, as follows 
\begin{eqnarray}
\label{eq:priorc}
\tilde\pi_0(\cbold|\thetabold,\tau, \lambda) &\propto&  
\exp\bigg\{-\frac{\lambda}{2}\sum_{i = 1}^{I}\int_{t_{1}+\tau}^{t_{{\color{black}\max}}}\bigg[\frac{dx_{i}(s)}{ds}-g_i(\xbold(s), \xbold(s-\tau)|\thetabold) \bigg]^{2}ds  \bigg\},  \nonumber \\
&=&  \exp\bigg\{-\frac{\lambda}{2}\sum_{i = 1}^{I}\int_{t_{1}+\tau}^{t_{{\color{black}\max}}}\bigg[\frac{d\Phibold_i(s)'}{ds} \cbold_{i}-g_{i}({\color{black} \Phibold(s)' \cbold, \Phibold(s-\tau)' \cbold} |\thetabold) \bigg]^{2}ds \bigg \},
\end{eqnarray}
where  $\Phibold(s) = \text{Diag}(\Phibold_1(s), \Phibold_2(s), \ldots, \Phibold_I(s))$  is a $\sum_i L_i\times I$-dimensional block diagonal matrix such that the $i$-th column contains $\Phibold_i(x)$ in the appropriate diagonal block and 0 elsewhere. This prior distribution measures how well the estimated DE variables $\xbold(t)$ satisfy the DE system {\color{black} defined on $[t_1, t_{\color{black} \max}]$}. It is based on treating a penalty term proposed in  \cite{ramsay2007parameter} as a prior similar to how \cite{berry2002bayesian} incorporated a penalty as a prior. The smoothing parameter $\lambda$ controls the trade-off between fit to the data and fidelity to the DE model.
  Details on selecting a proper $\lambda$ will be discussed in \emph{Section} \ref{sec: lambda}.

In the  Bayesian framework, we need to assign appropriate priors for 
model  parameters $\thetabold$, $\tau$, $\sigma_{i}^{2}$, $i = 1, \ldots, I$. The following priors are specified:
 \begin{eqnarray}
  \label{priors1}
\thetabold &\sim& \text{MVN}(\boldsymbol{0}_D,\sigma_{\theta}^2\Ibold_D), \\
  \label{priors2}
 \tau &\sim& \text{Unif}(t_{1}, t_{{\color{black}\max}}), \\
  \label{priors4} 
 \sigma_i^2 &\sim& \text{IG}(g_0,h_0),~ i=1,\ldots,I, 
 \end{eqnarray}
where $\sigma_{\theta}^{2}$, $g_{0}$, and $h_{0}$ are the hyper-parameters in prior distributions, and $D$ is the dimension of the vector $\thetabold$. The vector of all zeros is represented by $\boldsymbol{0}$, and
$\Ibold$ is an identity matrix. Their subscripts denote the vector/matrix dimension.

\subsection{The choice of $\lambda$}
\label{sec: lambda}

The tuning parameter $\lambda$ is important in balancing between fit to the data and fidelity to the DE model. A small  value of $\lambda$ does not impose much information about the DE fitting.  If $\lambda\to 0$, we end up fitting least squares for spline  coefficients with the data. If we choose a large value of $\lambda$,  the prior information of {\color{black} the} DE system is too strong and not much information about the data is taken into consideration. Hence, it is crucial to choose a proper value of $\lambda$ to balance the DE fitting and data information. 

One approach to choose $\lambda$ is through cross-validation \citep{wang2012estimating, reiss2009smoothing} from a range of reasonable choices of $\lambda$. However, this approach is infeasible in a Bayesian framework as it significantly increases the computational cost. We propose to treat $\lambda$ as an unknown parameter by specifying a prior distribution on $\lambda$  and estimating its posterior distribution through a Bayesian method. This idea is adapted from \cite{berry2002bayesian}, in which they automatically select a smoothing parameter for splines. We choose the prior distribution for the smoothing parameter to be $\text{Gamma}(a_{\lambda}, b_{\lambda})$.

\subsection{Posterior distribution of DE model} \label{sec:post2}
The likelihood function is
\begin{eqnarray*}
&&p(\ybold|\tau, \thetabold, \cbold, \boldsymbol{\sigma})\propto \left(\prod_{\textcolor{black}{i\in \mathcal{I}_0}}\prod_{j=1}^{{\color{black} J_i}}\sigma_{i}^{2}\right)^{-1/2}\exp\bigg\{ -   \sum_{\textcolor{black}{i\in \mathcal{I}_0}}  \bigg(\sum_{j=1}^{{\color{black} J_i}}\frac{(y_{ij} -  \Phibold_i(t_{ij})' \cbold_{i})^{2}}{2\sigma_{i}^{2}} \bigg )\bigg\}.
\end{eqnarray*}

We introduce a new notation $\betabold = (\thetabold', \tau, \cbold', \boldsymbol{\sigma}', \lambda)'$ to denote all the parameters of interest.  Let $\tilde\pi_{0}(\betabold)$ denote the prior distribution, which is specified in Equations {\color{black} (\ref{eq:priorc})} to (\ref{priors4}), and $\text{Gamma}(a_{\lambda}, b_{\lambda})$ for $\lambda$.  We are interested in the posterior distribution for $\betabold$
\[
\pi(\betabold) \propto {\gamma(\betabold)} = {p(\ybold|\tau, \thetabold, \cbold, \boldsymbol{\sigma}) \tilde\pi_{0}(\cbold|\thetabold,\tau, \lambda) \tilde\pi_{0}(\thetabold) \tilde\pi_{0}(\tau) \tilde\pi_0(\boldsymbol{\sigma}) \tilde\pi_{0}(\lambda)}.
\]
Here $\gamma(\betabold)=\tilde\pi_{0}(\betabold)p(\ybold|\betabold)$ is the unnormalized posterior distribution of $\betabold$ and can be written as 
{\small
\begin{eqnarray*}
&&\gamma( \betabold)  \propto  \\ &&( \prod_{\textcolor{black}{i\in \mathcal{I}_0}} \prod_{j=1}^{{\color{black} J_i}}\sigma_{i}^{2})^{-1/2}\exp\bigg\{ -  
 \sum_{\textcolor{black}{i\in \mathcal{I}_0}} \sum_{j=1}^{{\color{black} J_i}}\frac{(y_{ij} -  \Phibold_i(t_{ij})' \cbold_{i})^{2}}{2\sigma_{i}^{2}} - \frac{\lambda}{2} \sum_{i=1}^{I}\int_{t_{1}+\tau}^{t_{{\color{black}\max}}}\bigg[\frac{d\Phibold_i(s)'}{ds} \cbold_{i}-g_{i}({\color{black} \Phibold(s)' \cbold, \Phibold(s-\tau)' \cbold} |\thetabold) \bigg]^{2}ds  \bigg\}\nonumber\\
&&\cdot (\prod_{\textcolor{black}{i \in \mathcal{I}_0}} \sigma_{i}^{2} )^{-g_{0}-1}\exp\bigg\{-\sum_{{\textcolor{black}{i \in \mathcal{I}_0}} } \frac{h_{0}}{\sigma_{i}^{2}} \bigg\}
\cdot\exp\bigg\{-\frac{\thetabold'\thetabold}{\sigma_{\theta}^{2}}\bigg\}\cdot \lambda^{a_{\lambda}-1}\exp (- b_{\lambda}\lambda).
\end{eqnarray*}

The integral 
\[
\Rbold_i = \int_{t_{1}+\tau}^{t_{{\color{black}\max}}}\bigg[\frac{d\Phibold_i(s)'}{ds} \cbold_{i}-g_{i}({\color{black} \Phibold(s)' \cbold, \Phibold(s-\tau)' \cbold} |\thetabold) \bigg]^{2}ds 
\]
usually does not have a closed-form expression.  However, it can be evaluated by numerical quadrature approximation.  Let $\eta_{i0} = t_1+\tau < \eta_{i1} <  \eta_{i2} < \cdots < \eta_{i\zeta_i}< t_{{\color{black}\max}} = \eta_{i(\zeta_i+1)}$ denote the knots placed within $[t_1+\tau, t_{{\color{black}\max}}]$ for $i$-th  DE function, 
we approximate the integral by using the composite Simpson's rule \citep{burden2001numerical}
\begin{displaymath}
\Rbold_i {\color{black}\approx} \sum_{l_i = 0}^{\zeta_i}\sum_{m=1}^{M}v_{l_i m}\cdot\bigg(\bigg[\frac{d\Phibold_i(s)'}{ds} \cbold_{i}-g_{i}({\color{black} \Phibold(s)' \cbold , \Phibold(s-\tau)' \cbold}|\thetabold) \bigg]^{2}\bigg|_{s = \xi_{il_im}}\bigg),
\end{displaymath}
where $M$ is the number of quadrature points, {\color{black} $\zeta_i$ is the number of knots used for the $i$-th DE function}, $\xi_{il_im}$ is the $m$-th quadrature point in $[\eta_{il_i}, \eta_{i(l_{i}+1)}]$, and $v_{l_im}$ is the corresponding quadrature weight.

\section{Methodology}
\label{sec:3}

One classical methodology for Bayesian inference of nonlinear DE parameters is Markov chain Monte Carlo (MCMC). In MCMC, we construct an ergodic Markov chain which admits the normalized posterior as its stationary distribution. 
If we run the chain long enough, convergence to the posterior is  guaranteed. We show the details of this method in  {\it Supplementary Section {\color{black} 2.1}}.

However, MCMC (more specifically, the Metropolis{\color{black} -}Hastings (MH) algorithm) is inefficient for estimating parameters of nonlinear DEs for several reasons. First, the posterior surface is extremely sensitive to DE parameters $\thetabold$. There may exist isolated modes in the posterior distribution. The posterior may change quite a bit even with a tiny change in the parameter value. 
Second, the computation of the likelihood function involves numerically solving nonlinear DEs, which is computationally expensive. Third, the convergence of MCMC is generally difficult to assess.

\subsection{An annealed sequential Monte Carlo for Bayesian DE  inference}
\label{sec:asmc}

To better cope with the {\color{black} inadequateness} of MCMC, we propose a sequential Monte Carlo (SMC) algorithm in the  SMC  framework of \cite{del2006sequential} for the \emph{static} setting for Bayesian DEs. This special case of  SMC is a  generic method to approximate a sequence of intermediate probability distributions $\{\pi_{r}(\betabold)\}_{0\leq r\leq R}$ 
defined on a common measurable space $(E, \mathcal{E})$. This method is different from the standard SMC algorithm \citep{doucet2000sequential, doucet2001introduction}, as the sequence of intermediate probability distributions $\{\pi_{r}(\betabold)\}_{0\leq r\leq R}$ in standard SMC methods are generally defined on measurable spaces with  increasing dimension.  

The SMC algorithm in the static setting approximates the target distribution $\pi(\betabold)$ in $R$ steps.
We are interested in sequentially sampling from the distributions $\{\pi_{r}(\betabold)\}_{0\leq r\leq R}$. For example, we first approximate $\pi_{0}(\betabold)$, then approximate $\pi_{1}(\betabold)$ and so on. The subscript $r$ denotes the index of intermediate probability distributions. The last intermediate target distribution $\pi_{R}(\betabold)$ is $\pi(\betabold)$. We let 
 $\pi_{r}(\betabold) \propto \gamma_{r}(\betabold)$,
where $\gamma_{r}$ can be evaluated pointwise. 
There are several reasons to use multiple distributions in SMC. First, in online problems, data arrive sequentially and we aim to do inference sequentially in time. The distribution $\pi_r(\betabold)$ is then the unnormalized posterior distribution conditioning on the first $r$ batches of data. Second, as in this work, a sequence of intermediate target distributions is introduced to 
facilitate the exploration of the state space.
At each step $r$, we use a collection of $K$ samples to represent $\pi_{r}(\betabold)$, denoted by $\{\betabold_{r}^{(k)}\}_{k = 1}^K$. Each of these $K$ samples is called a particle. There is a positive weight associated with each particle $\betabold_{r}^{(k)}$. We use $w_r^{(k)}$ to denote the unnormalized weight of $\betabold_{r}^{(k)}$ and use $W_{r}^{(k)}$ to be the corresponding normalized weight. 
From iteration $r$ to $r+1$, 
we move particles from $\{\betabold_{r}^{(k)}\}_{k = 1}^K$ to $\{\betabold_{r+1}^{(k)}\}_{k = 1}^K$ using a Markov kernel denoted $T_{r+1}(\betabold_{r}^{(k)}, \betabold_{r+1}^{(k)})$. One typical approach in the SMC framework for the static setting is to select $T_{r+1}(\betabold_{r}^{(k)}, \betabold_{r+1}^{(k)})$ to be a $\pi_{r+1}$-invariant MCMC kernel; this will be detailed later in the paper. Then we compensate for the difference between the particles $\betabold_{r+1}^{(k)}$ proposed from $\{T_{r+1}(\betabold_{r}^{(k)}, \betabold_{r+1}^{(k)})\}_{k = 1}^K$ and $\pi_{r}(\betabold)$ by the updated weights $W_{r+1}^{(k)}$.  To get $W_{r+1}^{(k)}$, we first compute the incremental importance weight 
\begin{eqnarray*}
\tilde{w}_{r+1}^{(k)} &=& \frac{\gamma_{r+1}(\betabold_{r+1}^{(k)}) L_{r}( \betabold_{r+1}^{(k)}, \betabold_{r}^{(k)}) }{\gamma_{r}(\betabold_{r}^{(k)}) T_{r+1}(\betabold_{r}^{(k)}, \betabold_{r+1}^{(k)}) }, 
\end{eqnarray*}
where $L_{r}( \betabold_{r+1}^{(k)}, \betabold_{r}^{(k)})$ is an artificial backward kernel \citep{del2006sequential, del2012adaptive}, denoting the probability of moving from  $\betabold_{r+1}^{(k)}$  to $\betabold_{r}^{(k)}$. Then we calculate the unnormalized weight by using the previous unnormalized weight and the incremental importance weight as follows
\begin{eqnarray*}
w_{r+1}^{(k)} &=& w_{r}^{(k)}\cdot \tilde{w}_{r+1}^{(k)}.
\end{eqnarray*}
The normalized weights $W_{r+1}^{(k)}$ are obtained by 
$W_{ r+1}^{(k)} = w_{r+1}^{(k)} /(\sum_{k=1}^{K}w_{r+1}^{(k)})$.

The selection of the backward kernel $L_{r}( \betabold_{r+1}^{(k)}, \betabold_{r}^{(k)})$ is important as it will impact the variance of $\{W_{r+1}^{(k)}\}_{k=1}^K$. We refer readers to \cite{del2006sequential} for a more detailed discussion of this artificial backward kernel.
A convenient backward  Markov kernel that allows an easy evaluation of the importance weight is
\begin{align*}\label{eqn:backwardKernel}
L_{r}( \betabold_{r+1}^{(k)}, \betabold_{r}^{(k)})
=\frac{\pi_{r+1}(\betabold_{r}^{(k)}){T}_{r+1}(\betabold_{r}^{(k)},\betabold_{r+1}^{(k)})}{\pi_{r+1}(\betabold_{r+1}^{(k)})}.
\end{align*}
With this backward kernel, the weight update function $\tilde{w}_{r+1}^{(k)}$ becomes
\begin{eqnarray*}
	\tilde{w}_{r+1}^{(k)} &=& \frac{\gamma_{r+1}(\betabold_{r+1}^{(k)}) L_{r}( \betabold_{r+1}^{(k)}, \betabold_{r}^{(k)}) }{\gamma_{r}(\betabold_{r}^{(k)}) T_{r+1}(\betabold_{r}^{(k)}, \betabold_{r+1}^{(k)}) }  \\ &=& 
	\frac{\gamma_{r+1}(\betabold_{r+1}^{(k)})}{\gamma_{r}(\betabold_{r}^{(k)})} \cdot
	\frac{\pi_{r+1}(\betabold_{r}^{(k)}){T}_{r+1}(\betabold_{r}^{(k)},\betabold_{r+1}^{(k)})}{\pi_{r+1}(\betabold_{r+1}^{(k)})} \cdot \frac{1}{T_{r+1}(\betabold_{r}^{(k)}, \betabold_{r+1}^{(k)})} \\ 	&=& \frac{\gamma_{r+1}(\betabold_{r}^{(k)})}{\gamma_{r}(\betabold_{r}^{(k)})}.  
\end{eqnarray*}
Thus, we do not require pointwise evaluation of the forward kernel $T_{r+1}(\betabold_{r}^{(k)}, \betabold_{r+1}^{(k)})$ and the backward kernel $L_{r}( \betabold_{r+1}^{(k)}, \betabold_{r}^{(k)})$ to compute the weight {\color{black} update} function.

In this article, we propose a sequence of annealing intermediate target distributions \citep{neal2001annealed, wang2018annealed} $\{\pi_{r}(\betabold)\}_{0\leq r\leq R}$ to facilitate the exploration of posterior space, such that 
\[
\pi_{r}(\betabold) \propto  \gamma_{r}(\betabold)= {[p(\ybold|\betabold)} \tilde\pi_{0}(\betabold)]^{\alpha_{r}} \rho(\betabold)^{1-{\alpha_{r}}},
\]
where $\rho(\betabold)$ is a reference distribution \citep{fan2011choosing}, and $0 = \alpha_{0} < \alpha_{1} < \cdots < \alpha_{R-1} < \alpha_{R} = 1$ is the sequence of annealing parameters.  When  $\alpha_{0}=0$, the first distribution is the reference distribution $\rho(\betabold)$; when  $\alpha_{R} = 1$, the last distribution is our target distribution, the posterior distribution of $\betabold$. 

\textcolor{black}{The reference distributions should be easy to sample from and ideally they are close to the modes of the target distribution. 
Since we can easily sample from all the prior distributions except for the prior of $\cbold$, we use the same reference distribution as the prior distribution for all parameters except for $\cbold$. 
}

In this case, 
\[
\pi_{r}(\betabold) \propto  \gamma_{r}(\betabold) = {[p(\ybold|\betabold)} \tilde\pi_{0}(\cbold|\thetabold,\tau, \lambda)]^{\alpha_{r}} \rho(\cbold)^{1-{\alpha_{r}}}\tilde\pi_{0}(\thetabold) \tilde\pi_{0}(\tau) \tilde\pi_{0}(\lambda).
\]

{\color{black} We specify the following reference distribution for $\cbold$ based on its MLE:  
 \begin{eqnarray}
  \label{priors3}
\cbold_{i} &\sim& \text{MVN}(\hat{\cbold}_{i}, \sigma_{c}^2\Ibold_{L_{i}}), ~ i=1,\ldots,I,
 \end{eqnarray}
where $\hat{\cbold}_{i}$ is the MLE of $\cbold_{i}$ by maximizing $p(\ybold|\betabold)\tilde\pi_{0}(\cbold|\thetabold,\tau, \lambda)$ with respect to $\tau$, $\thetabold$, $\cbold$, and $\boldsymbol{\sigma}$ given $\lambda$, $\sigma_{c}^{2}$ is the hyper-parameter in the reference distribution. }

If there are isolated modes in $\pi(\betabold)$, MCMC may get stuck in one of the modes which is close to
the initial value. A sequence of intermediate distributions is introduced to avoid this. With a small annealing parameter $\alpha_{r}$, the intermediate distribution surface is flat, which makes  MCMC samples move easily between modes. The intermediate distribution with a higher value of annealing parameter is  closer to the true posterior. The samples move closer to the target posterior distribution if we increase $\alpha_{r}$. 
One simple choice of annealing parameters is to equally put parameters across $[0,1]$, such that $\alpha_{0} = 0$, $\alpha_{1} = 1/R, \alpha_{2} = 2/R, \ldots, \alpha_{R-1} = (R-1)/R, \alpha_{R} = 1$.

We now introduce an SMC algorithm with a defined sequence of intermediate targets. 
First, we initialize particles $\{\betabold_{0}^{(k)}\}_{k = 1}^{K}$. 
At each step $r-1$, we keep a list of $K$ particles $\{\betabold_{r-1}^{(k)}\}_{k = 1}^K$ in memory. We let $\{\tilde{\betabold}_{r-1}^{(k)}\}_{k = 1}^{K}$ denote particles after  the resampling step (see {\it Step 3}). We iterate between the following three steps to obtain the approximated intermediate  distributions
\begin{eqnarray*}
\hat{\pi}_{r}(\betabold) = \sum_{k = 1}^{K}W_{r}^{(k)}\cdot \delta_{\betabold_{r}^{(k)}}(\betabold), (r = 1, \ldots, R).
\end{eqnarray*}

{\it Step 1. }
We compute the weight function for particles at  iteration $r$ with
\begin{eqnarray}
\label{eq: weights}
W_{r}^{(k)}\propto w_{r}^{(k)} = w_{r-1}^{(k)}\cdot\frac{\gamma_{r}(\tilde\betabold_{r-1}^{(k)})}{\gamma_{r-1}(\tilde\betabold_{r-1}^{(k)})} = w_{r-1}^{(k)} \left( \frac{p(\ybold|\tilde\betabold_{r-1}^{(k)}) \tilde\pi_0(\tilde\betabold_{r-1}^{(k)})}{\rho(\tilde\betabold_{r-1}^{(k)})}\right)^{\alpha_{r} - \alpha_{r-1}}.
\end{eqnarray}
Note that the weight update function for particles at the $r$-th iteration only depends on particles at the $(r-1)$-th iteration, which is different from the standard SMC algorithm \citep{doucet2000sequential, doucet2001introduction}.

{\it Step 2. }
We propagate new samples $\{\betabold_{r}^{(k)}\}_{k = 1}^K$ via $\pi_{r}$-invariant MCMC moves, $\{\betabold_{r}^{(k)} \sim T_{r}(\tilde\betabold_{r-1}^{(k)}, \cdot)\}_{k = 1}^K$. The full conditional posterior distributions, $\pi_{r}(\sigma_{i}^{2}|\cbold_{i})$, $\pi_{r}( \tau |\cbold, \thetabold, \lambda)$, $\pi_{r}( \thetabold |\cbold,\tau, \lambda)$ and $\pi_{r}( \cbold_{i} |\tau, \thetabold, \boldsymbol{\sigma}, \cbold_{-i}, \lambda)$ are described in {\it Supplementary Section 1}.

{\it Step 3. } We conduct a resampling step to prune particles with small weights. 
The particles after the resampling step are denoted by $\{\tilde\betabold_{r}^{(k)}\}_{k = 1}^K$, and  all particles are equally weighted. The simplest resampling method is multinomial resampling based on the normalized particle weights. However, advanced resampling schemes such as stratified resampling {\color{black}\citep{kitagawa1996monte, hol2006resampling}} or residual resampling \citep{douc2005comparison} are preferable to multinomial resampling, since multinomial resampling will create more variance for the SMC estimator when compared with advanced resampling algorithms. In our numerical experiments, we use systematic resampling \citep{carpenter1999improved}.

It is not recommended to conduct resampling at every iteration as resampling will {\color{black} create} additional variation in the estimator \citep{chopin2004central}.
Our resampling scheme is typically triggered when the relative effective sample size (rESS) falls below a given thresholds $\varsigma$. 
 The effective sample size (ESS) at iteration $r$ can be computed by 
\[
\text{ESS}_{r}^{(K)} = \frac{1}{\sum_{k = 1}^{K}(W_{r}^{(k)})^{2}}.
\]
$\text{ESS}_{r}^{(K)}$ denotes the number of ``perfect" samples used to approximate the intermediate distribution $\pi_{r}$. Effective sample size takes value between $1$ and $K$. It takes value $K$ if all particles are equally weighted, and it takes value close to $1$ if one of the particles has a much larger weight than the others. The rESS normalizes the ESS to be between zero and one. The rESS at iteration $r$ can be computed by $\text{rESS}_{r}^{(K)} = \text{ESS}_{r}^{(K)}/K$. If we never conduct resampling, the annealed SMC algorithm degenerates to the annealed importance sampling \citep{neal2001annealed}.

After conducting the annealed SMC algorithm, we obtain a list of weighted samples to empirically represent the posterior distribution $\pi(\betabold)$,
\[
\hat{\pi}(\betabold) = \sum_{k = 1}^{K}W_{R}^{(k)}\cdot \delta_{\betabold_{R}^{(k)}}(\betabold).
\]

\subsection{Properties of the annealed SMC algorithm}
\label{sec:prop}

We discuss some properties of our annealed SMC method. {\color{black} Note that the general SMC algorithm proposed by \cite{del2006sequential} includes our annealed SMC as a special case. Hence {\em consistency} and {\em asymptotic normality} properties can be generated from \cite{del2006sequential}. We summarize these properties in following propositions.} 
First, our annealed SMC method can provide a consistent representation of the intermediate target posterior distributions.

\noindent {\it Proposition 1. } The annealed SMC method provides asymptotically consistent estimates. We have
\[
\sum_{k = 1}^{K}W_{r}^{(k)}\psi(\betabold_{r}^{(k)}) \to \int \pi_{ r}(\betabold)\psi(\betabold)\text{d}\betabold ~~\text{as}~~K\to \infty,
\]
where the convergence is in $L^{2}$ norm, and $\psi$ is a target function that satisfies regularity conditions, for example $\psi$ is bounded. \cite{del2004feynman} and \cite{chopin2004central} discussed more general conditions which include the case of our annealed SMC algorithm.

The central limit theorem shown below can be used to assess the total variance of Monte Carlo estimators. 

\noindent {\it Proposition 2. } Under the integrability conditions given in Theorem $1$ of \cite{chopin2004central}, or \cite{del2004feynman}, pages $300{\color{black}\textendash}306$ in Section $9.4$, {\color{black} when multinomial resampling is performed at each iteration,}
\[
K^{1/2}\bigg[ \sum_{k=1}^K W_{r}^{(k)} \psi(\betabold_{r}^{(k)})  - \int \pi_r(\betabold) \psi(\betabold) d \betabold\bigg]\to N(0, \sigma_{r}^{2}(\psi)) ~~\text{as}~~K\to \infty,
\]
where the convergence is in distribution. The form of asymptotic variance $\sigma_{r}^{2}(\psi)$ depends on  the Markov kernel $T_{r}$, and the artificial backward kernel $L_{r}$. We refer readers to \cite{del2006sequential} for details of this asymptotic variance. 

Note that {\it Propositions 1{\color{black}\textendash}2} hold if the integral in {\color{black} $\Rbold_{i}$} can be computed exactly. These propositions do not hold exactly if we numerically approximate the integral due to the error being introduced. 

In addition, the annealed SMC algorithm can be easily parallelized, by allocating particles across different cores \citep{del2006sequential, wang2018annealed}.

\subsection{Adaptive annealing parameter scheme in SMC}
\label{sec:adaptive}
In the annealed SMC algorithm, one challenge is to properly select the sequence  of annealing parameters. If we choose $\alpha_{0} = 0$ and $\alpha_{1} = 1$, the annealed SMC {\color{black} sampler} degenerates to importance sampling. 
A large number of annealing parameters improves the performance of algorithm, but it will be computationally more intensive. If we select an insufficient number of annealing parameters or an improper annealing scheme, the algorithm may collapse. 
We propose an adaptive annealing parameter scheme based on the seminal work of \cite{del2012adaptive, zhou2016toward, wang2018annealed}. Note that the weight function (Equation \ref{eq: weights}) for iteration $r$ only depends on particles of  the $(r-1)$-th iteration, and the difference between two successive annealing parameters $\alpha_{r} - \alpha_{r-1}$. This indicates that we can ``manipulate" $\tilde{w}_{r}^{(k)}$ by changing the annealing parameter $\alpha_{r}$. If $\alpha_{r}$ is close to $\alpha_{r-1}$, the incremental weight function $\tilde{w}_{r}^{(k)}$ is close to $1$, and the variance of $\tilde{w}_{r}^{(k)}$ is smaller than it would be if we {\color{black} choose} a larger value of $\alpha_{r}$. This provides the intuition that we are able to control the discrepancy between two successive intermediate target distributions by manipulating $\alpha_{r}$.

In this article, we use the relative conditional effective sample size (rCESS) \citep{zhou2016toward} to measure the discrepancy between two successive intermediate targets. The rCESS normalizes the conditional effective sample size (CESS) to be between zero and one. The rCESS is defined as
\begin{eqnarray*}
\text{rCESS}_{r}(W_{r-1}^{(\cdot)}, \tilde{w}_{r}^{(\cdot)})  = \frac{\big(\sum_{k = 1}^{K}W_{r-1}^{(k)}\tilde{w}_{r}^{(k)}\big)^{2}}{\sum_{k = 1}^{K}W_{r-1}^{(k)}\big(\tilde{w}_{r}^{(k)}\big)^{2}},
\end{eqnarray*}
which takes a value between $1/K$ and $1$. 
The rCESS is equal to the relative ESS if we conduct resampling at every SMC iteration.  
Using the fact that $\tilde{w}_{r}^{(k)} = [p(\ybold|\betabold_{r-1}^{(k)})\pi_0(\betabold_{r-1}^{(k)})/\rho(\betabold_{r-1}^{(k)})]^{\alpha_{r} - \alpha_{r-1}}$, $\text{rCESS}_{r}$ is a decreasing function of $\alpha_{r}$, where $\alpha_{r}\in (\alpha_{r-1}, 1]$. We control rCESS over iterations by selecting the annealing parameter $\alpha$ such that 
$$f(\alpha) = \text{rCESS}\left(W_{r-1}^{(\cdot)}, \tilde{w}_{r}^{(\cdot)}\right) = \phi,$$
where $\phi$ is a value between $0$ and $1$. A small value of $\phi$ will lead to a high value of $\alpha_{r}$, while a large value of $\phi$ will lead to a low value of $\alpha_{r}$. It is impossible to obtain a closed-form solution of $\alpha^*$ by solving $f(\alpha) = \phi$, but we are able to use a bisection method to solve this one-dimensional search problem. The search interval of $\alpha$ is $(\alpha_{r-1},1]$.  
By using $f(\alpha_{r-1}) - \phi >0$, $f(1) - \phi < 0$ (in case $f(1) \geq \phi $, we set $\alpha_{r} = 1$), and the continuous property of $f(\alpha) - \phi$, the solution $\alpha^*$ of $f(\alpha) = \phi$ is guaranteed. Algorithm \ref{algo:adapt} provides a detailed description for the SMC algorithm. 

\begin{algorithm}
   \caption{\bf{An SMC algorithm of Bayesian inference for parameters in DEs}}
  \label{algo:adapt}
{\fontsize{10pt}{10pt}\selectfont
\begin{algorithmic}[1]
  \State {\bfseries Inputs:} (a) The prior distribution $\tilde\pi_{0}(\cdot)$ and the reference distribution $\rho(\cdot)$ over model parameters $\betabold$, where $\betabold = (\thetabold', \tau,  \cbold', \boldsymbol{\sigma}', \lambda)'$;  (b) relative CESS {\color{black} threshold} $\phi$; (c) resampling threshold $\varsigma$. 

   \State {\bfseries Outputs:}   Posterior approximation, $\hat{\pi}(\betabold) = \sum_{k = 1}^{K}W_{R}^{(k)}\cdot \delta_{\betabold_{R}^{(k)}}(\betabold)$.

	\State  Initialize the SMC iteration index and annealing parameter: $r \leftarrow 0$, $\alpha_0 \leftarrow 0$.  
   \For{$k \in \{1, 2, \dots, K\}$}
    \State  Initialize particles with independent samples from the reference distribution:
    $$\betabold_{0}^{(k)} \leftarrow ({\thetabold_{0}^{(k)}}', \tau_{0}^{(k)}, {\cbold_{0}^{(k)}}', {\boldsymbol{\sigma}_{0}^{(k)}}', \lambda_{0}^{(k)})'.$$ 
	\State  Initialize weights to unity: $w_{0}^{(k)} \leftarrow 1$, $W_{0}^{(k)} \leftarrow 1/K$. 
\EndFor    
 \For{$r\in \{1, 2, \dots\}$}
    \State Compute annealing parameter $\alpha_{r}$ using {\color{black} a} bisection method with $$f(\alpha) = \text{rCESS}\left(W_{r-1}^{(\cdot)}, \left( \frac{p(\ybold|\tilde\betabold_{r-1}^{(\cdot)}) \tilde\pi_0(\tilde\betabold_{r-1}^{(\cdot)})}{\rho(\tilde\betabold_{r-1}^{(\cdot)})}\right)^{\alpha_{r} - \alpha_{r-1}} \right) =\phi.$$ 
    \For{$k \in \{1, \dots, K\}$}
      \State \label{step:adapt-weigh} Compute unnormalized weights for $\betabold_{r}^{(k)}$: $w_{r}^{(k)} =  w_{r-1}^{(k)}\cdot  \left( \frac{p(\ybold|\tilde\betabold_{r-1}^{(k)}) \tilde\pi_0(\tilde\betabold_{r-1}^{(k)})}{\rho(\tilde\betabold_{r-1}^{(k)})}\right)^{\alpha_{r} - \alpha_{r-1}} $.
      \State Normalize weights: $W_{r}^{(k)}=w_{r}^{(k)} /(\sum_{k=1}^{K}w_{r}^{(k)})$.
        \State Sample particles $\betabold_{r}^{(k)}$ with one MCMC move admitting $\pi_r$ as  stationary/invariant distribution{\color{black}, using particles $\tilde{\beta}_{r-1}^{(k)}$ and the propagation step in {\it Supplementary Section 1}}. 
    \EndFor    
     \If{$\alpha_r = 1$} 
   \State \label{step:adapt-return} return the current particle population $\{(\betabold_{r}^{(k)}, W_{r}^{(k)})\}_{k = 1}^K$.
   \Else
  \If{$\text{rESS} < \varsigma$}  
		\State Resample the particles. 
    \For{$k \in \{1, \dots, K\}$}
    \State \label{step:adapt-reset-weights} Reset particle weights: $w_{r}^{(k)} = 1, W_{r}^{(k)} = 1/K$.
    \EndFor
    \Else
    \For{$k \in \{1, \dots, K\}$}
    \State $\tilde \betabold_{r}^{(k)} = \betabold_{r}^{(k)}$. 
    \EndFor
   \EndIf
      \EndIf
\EndFor 
 
\end{algorithmic}
}
\end{algorithm}

\section{Real Data Analysis}
\label{sec:real}
\begin{figure}[ht]
\center
\includegraphics[scale=0.8]{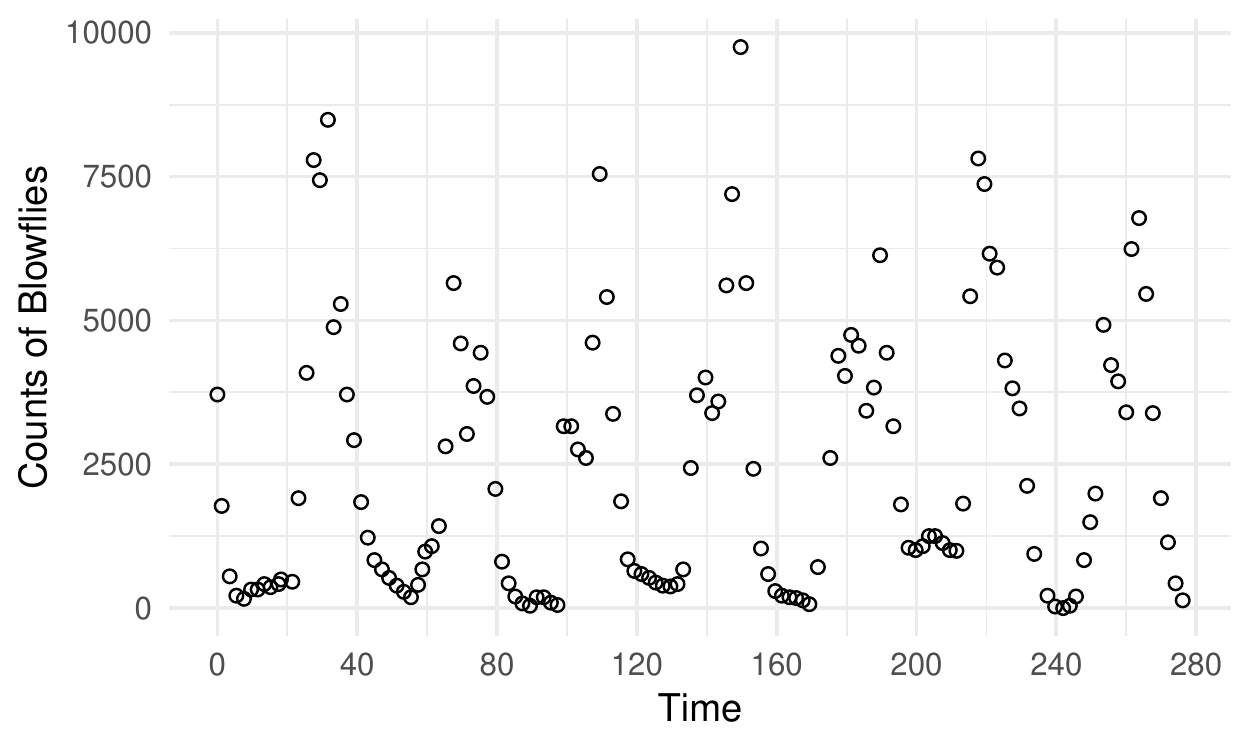}
\caption{Blowfly population in one experiment published in \cite{nicholson1954outline}; the time unit is one day.}
\label{fig:real}
\end{figure}

In the dynamic system of the blowfly population, resource limitation acts with a time
delay, roughly equal to the time for an egg to grow up to a pupa.
One classic experiment on the resource competition in laboratory populations of Australian sheep blowflies ({\it Lucilia cuprina}) is studied by \cite{nicholson1954outline}. The
blowflies were cultivated in a room with temperature maintained at 25\textdegree{}C. 
The population
of blowflies was measured every day for approximately one year.
 Figure \ref{fig:real} displays the counts of blowflies over time studied in \cite{nicholson1954outline}. The time unit is one day. 
The  oscillations displayed in the blowfly population are caused by the time lag between stimulus and reaction \citep{berezansky2010nicholson}. 
\cite{may1976models}
proposed to model the counts of blowflies with the following DDE model
\begin{eqnarray}
\frac{dx(t)}{dt} & = & \nu x(t)[1 - x(t-\tau)/(1000\cdot P)],
\end{eqnarray}
where $x(t)$ is the blowfly population, $\nu$ is the rate of increase of the blowfly population,
$P$ is a resource limitation parameter set by the supply of food, and $\tau$ is the time delay,
roughly equal to the time for an egg to grow up to a pupa. Our goal is to
estimate the initial value, $x(0)$, and the three parameters, $\nu$, $P$, and $\tau$, from the noisy Nicholson's blowfly data $y(t)$. 
The observed counts of blowflies $y(t)$ is assumed to be lognormal distributed with mean $x(t)$ and variance $\sigma^2$.

The counts of blowfly $x(t)$ is a positive function. Instead of modelling the constrained function $x(t)$ by a linear combination of cubic B-spline basis functions {\color{black}$W(t) = \Phibold(t)' \cbold$}, we transform $x(t) =e^{W(t)}$ and use B-spline basis functions to model the unconstrained function {\color{black}$W(t) = \Phibold(t)' \cbold$}. Equivalently, we solve the delay differential equation 
 \begin{eqnarray}
 \label{eq:12}
\frac{dW(t)}{dt} & = & \nu [1 - e^{W(t-\tau)}/(1000\cdot P)],
\end{eqnarray}
with noisy observations $\log y(t) \sim \text{N}(W(t), \sigma^{2})$.

We approximate the DDE solution using cubic B-splines with $34$ equally spaced interior knots over the time span. 
The total number of knots is equal to $36$. The total number of cubic B-spline functions is $L = 38$.  Selection of the number of basis functions is explored in  {\it  Section \ref{sec:tuningparameters}}. 
Our prior/reference distributions for parameter of interest $(\cbold, \nu, P, \tau, \sigma^2, \lambda)^{'}$ are 
 \begin{eqnarray*}
\nu &\sim& \text{N}(0,5^2)\text{I}(\nu>0), 
~~~P ~\sim~ \text{N}(0,5^2)\text{I}(P>0), ~~~\tau \sim \text{Unif}(0,50), \\
 \cbold &\sim& \text{MVN}({\color{black}\hat{\cbold}},100^2\Ibold_L),
~~~\sigma^2 \sim \text{IG}(1, 1), ~~~\lambda \sim \text{Gamma}(1, 1).
 \end{eqnarray*}
 In our adaptive SMC, we set the  rCESS threshold $\phi = 0.9$ and resampling threshold $\varsigma = 0.5$. The number of particles is $K = 500$. Under this setting, the number of SMC iteration $R = 227$.  Figure 2 in {\it Supplementary} shows the annealing parameter sequence $\alpha_{1:R}$ under the adaptive scheme. In {\it  Section \ref{sec:tuningparameters}}, we will compare the performance of our method using different values of $\phi$ and $K$.  Figure \ref{fig:realtrajectory} displays the estimated DDE trajectory. The left panel of Figure \ref{fig:realtrajectory} shows the estimated $W(t)$ and the $95\%$ {\color{black} pointwise} credible {\color{black} intervals}; the right panel of Figure \ref{fig:realtrajectory} shows $X(t) = e^{W(t)}$, in which the blue points are observed data. 
 
\begin{figure}[ht]
\center
\includegraphics[scale=0.7]{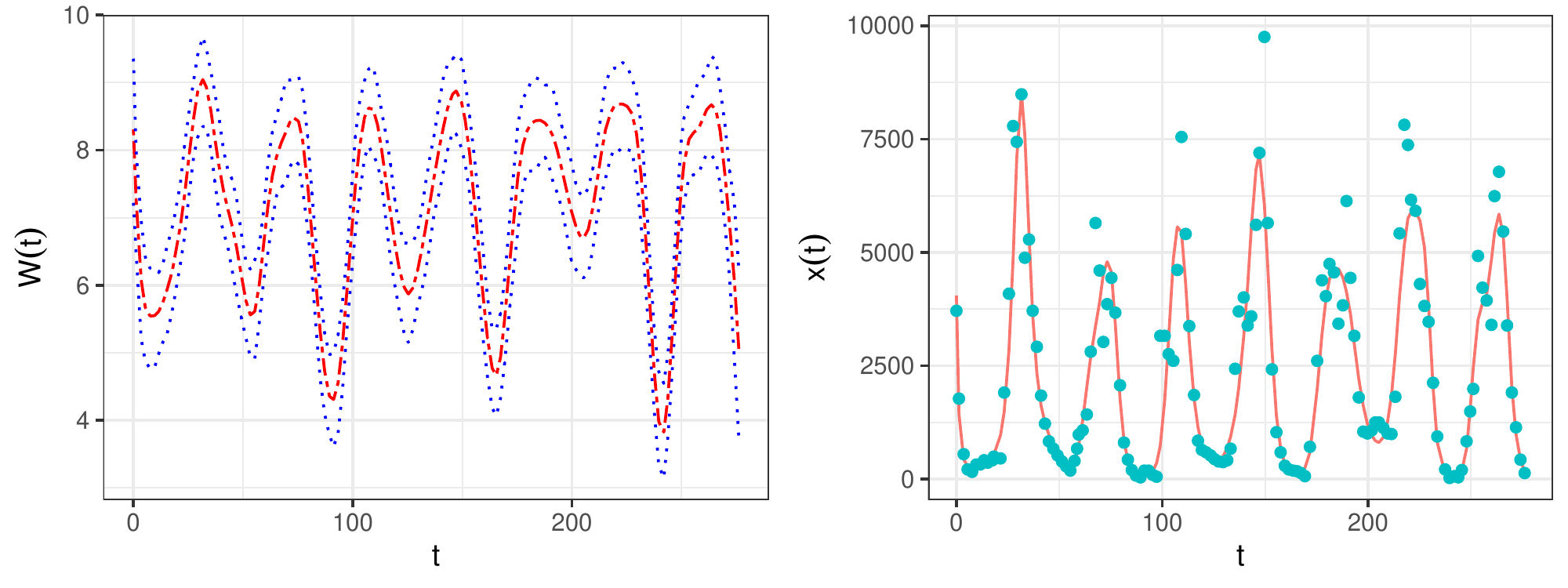}
\caption{Estimated posterior mean trajectory and 95\% {\color{black} pointwise credible intervals}  for the delay differential equation modelling the population dynamics of blowflies. }
\label{fig:realtrajectory}
\end{figure}

Table \ref{tab:2} displays the posterior means and the corresponding $95\%$ {\color{black} pointwise} credible intervals (CI) for DDE parameters in Equation (\ref{eq:12}). Note that our point estimates are similar to those obtained from  \cite{wang2012estimating}, in which  the same nonparametric function expressed using B-splines is estimated by maximizing the DDE-defined penalized likelihood function. However, the uncertainty of these parameters is significantly underestimated using their frequentist approach. In contrast, our Bayesian approach can provide more reasonable estimates for the parameter uncertainty. More concretely, we compare the estimates for the main parameter of interest, the delay parameter $\tau$, which can be interpreted as the time for an egg to grow up to a pupa.  From Table \ref{tab:2},  our posterior mean of $\tau$  and its 95\% CI is $8.368$ $(5.656, 9.916)$ while  
 the maximum likelihood estimate for $\tau$ is 8.781 and the standard error is 0.039  in \cite{wang2012estimating}.

Figure \ref{fig:realparticle} displays the pairwise scatter plots of the posterior samples of $\nu$, $\tau$, and $P$. We also calculate the correlation between posterior samples: $\corr(\nu, \tau) = 0.139$, $\corr(\nu, P) = 0.598$,  $\corr(P, \tau) = 0.008$. Recall that $\nu$  is the rate of increase of the blowfly population and $\tau$ is the time delay, roughly equal to the time for an egg to grow up to a pupa. The small  positive value of the correlation between $\tau$ and $\nu$ indicates that the blowfly population will increase 
 if eggs take their time to develop into pupae. The  parameter $P$ is related to a resource limitation.   The relatively large positive correlation between $\nu$ and $P$ can be easily understood: the blowfly population grows faster when there is a larger food supply.     The tiny positive value of the correlation between $\tau$ and $P$ implies that the amount of food supply has a small impact on the period of being a pupa.

\begin{table}[ht]
\centering
\caption{Posterior mean and corresponding $95\%$ credible interval (CI) for parameters in population dynamics of blowflies.}
\label{tab:2}
{
\begin{tabular}{rrrrrrr}
  \hline
  &$\nu$ & P & $\tau$ \\ 
  \hline
Mean &  0.18 & 2.37 &  8.37 \\ 
 (2.5\%, 97.5\%) & (0.07, 0.28)  &(1.31, 3.33) & (5.66, 9.92)    \\ 
   \hline
   & $W(0)$ & $\sigma^2$ & $\lambda$\\ 
    \hline
   Mean &  8.30 & 0.53 & 3.35\\ 
 (2.5\%, 97.5\%) & (7.22, 9.36)   &  (0.46, 0.61)& (1.63, 5.91)  \\ 
  \hline
\end{tabular}
}
\end{table}

\begin{figure}[ht]
\center
\includegraphics[scale=0.7]{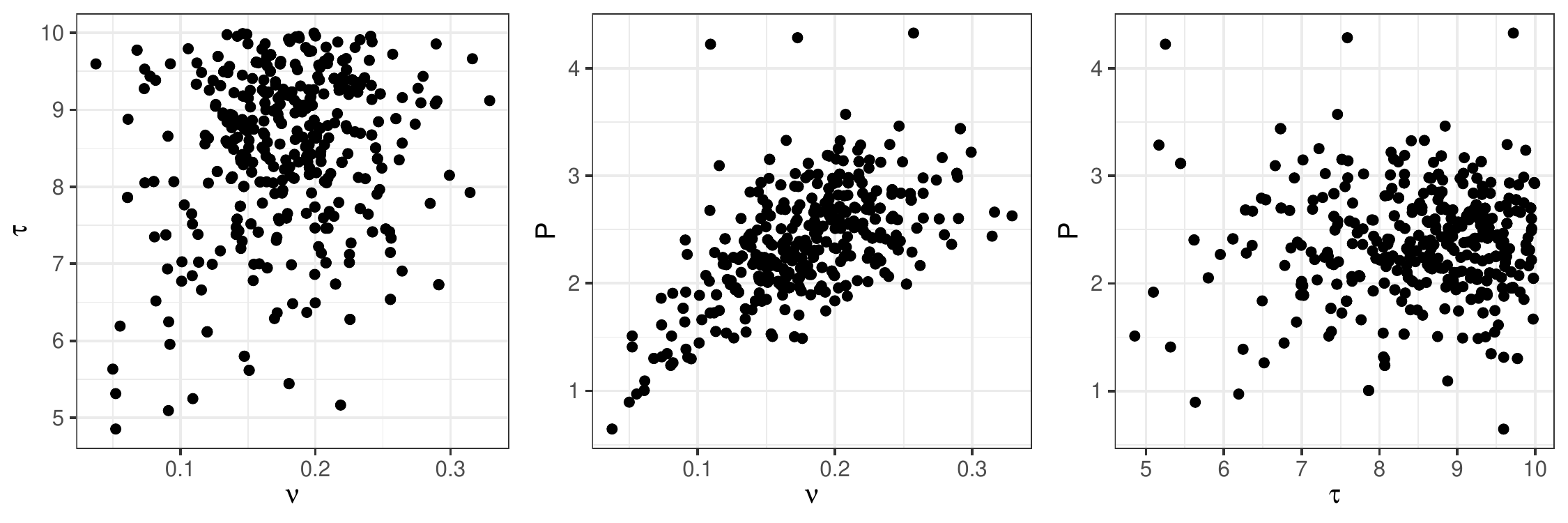}
\caption{Posterior samples of $\nu$, $\tau$, $P$  for DDE in Equation (\ref{eq:12}) estimated via SMC. We resample the particles at the last SMC iteration such that they are equally weighted. Correlation: $\corr(\nu, \tau) = 0.139$, $\corr(\nu, P) = 0.598$,  $\corr(P, \tau) = 0.008$.}
\label{fig:realparticle}
\end{figure}

\section{Simulation Study}
\label{sec:simulation}
We use simulation studies to demonstrate the effectiveness of our proposed model and method. The experiments include both ODE and DDE examples. We use the R package \emph{deSolve} \citep{soetaert2010solving} to simulate differential equations. 
\subsection{A nonlinear ordinary differential equation example}
\label{sec:sim1}
In this section, we use a nonlinear ODE example to illustrate the numerical behaviour of SMC algorithm.  We generate ODE trajectories according to the following ODE system,
\begin{eqnarray}
\label{eq: ode1}
\frac{dx_{1}(t)}{dt} &=& \frac{72}{36+x_{2}(t)} - \theta_{1},\nonumber\\
\frac{dx_{2}(t)}{dt} &=& \theta_{2}x_{1}(t) - 1,
\end{eqnarray}
where $\theta_{1} = 2$ and $\theta_{2} = 1$, and initial conditions $x_{1}(0) = 7$ and $x_{2}(0) = -10$. The observations $\ybold_{i}$ are simulated from a normal distribution with mean  $x_{i}(t|\thetabold)$ and variance $\sigma_{i}^{2}$, where $\sigma_{1} = 1$ and $\sigma_{2} = 3$. We generate $121$ observations for each ODE function, equally spaced within $[0, 60]$ (see Figure $3$ in {\it Supplementary}).
Under this setting, the posterior distribution of $\theta_{1}$ and $\theta_{2}$ will have multiple local modes (see Figure \ref{fig:graphandLL}).

We use cubic B-spline functions (see Figure $1$ in {\it Supplementary}) to represent ODE trajectories. We put equally spaced knots on each of eight observations. The total number of knots is $16$, including $14$ interior knots. The total number of cubic B-spline functions is ${\color{black} L_1 = L_2} = 18$. 
We select weak prior/reference distributions of $\betabold$ for the SMC algorithm,
 \begin{eqnarray*}
\theta_{1} &\sim& \text{N}(5,5^2), ~~~\theta_{2} ~\sim~ \text{N}(5,5^2), \\
 \cbold_{1} &\sim& \text{MVN}({\color{black}\hat{\cbold}_{1}},100^2\Ibold_{{\color{black} L_1}}),
 ~~~\cbold_{2} ~\sim~ \text{MVN}({\color{black}\hat{\cbold}_{2}},100^2\Ibold_{{\color{black} L_2}}),\\
 \sigma_1^2 &\sim& \text{IG}(1, 1), ~~~\sigma_2^2 ~\sim~ \text{IG}(1, 1), ~~~\lambda ~\sim~ \text{Gamma}(1, 1).
 \end{eqnarray*}

\subsubsection{Comparison of SMCs and MCMCs}
We first alter Equation (\ref{eq: ode1}) to produce a symmetric, bimodal
posterior for $\theta_{1}$, 
\begin{eqnarray}
\label{eq: ode2}
\frac{dx_{1}(t)}{dt} &=& \frac{72}{36+x_{2}(t)} - |\theta_{1}|,\nonumber\\
\frac{dx_{2}(t)}{dt} &=& \theta_{2}x_{1}(t) - 1.
\end{eqnarray}
We compare the performance of annealed SMC targeting $\pi(\betabold)$ (denoted SMC-spline) with the following three algorithms in terms of speed and estimation using Equation (\ref{eq: ode2}):   annealed SMC targeting $\pi(\thetabold, \tau, \xbold(0), \boldsymbol{\sigma}^{2})$ (SMC-deSolve), MCMC targeting $\pi(\betabold)$  (MCMC-spline), and MCMC targeting $\pi(\thetabold, \tau, \xbold(0), \boldsymbol{\sigma}^{2})$ (MCMC-deSolve).  {\it Supplementary Section 2}   details the three algorithms.

We simulate the ODE trajectories using the ``\emph{lsoda}" method in the R package \emph{deSolve} \citep{soetaert2010solving}. In SMC-spline, we set the rCESS threshold $\phi = 0.9$ and resampling threshold $\varsigma = 0.5$. The total number of particles we use is $K = 500$.  With given samples $\thetabold^{(n)}$, $\xbold(0)^{(n)}$ in the SMC-deSolve and MCMC-deSolve methods,  we use  the ``\emph{euler}" method in \emph{deSolve}  to solve ODEs to obtain $\xbold(t_{ij}|\thetabold^{(n)}, \xbold(0)^{(n)})$. We purposely use a different ODE solver to mimic the fact that no  data generation information  is available for real data. We select weak prior distributions for $\thetabold$ and $\xbold(0)$. 
In the {\color{black} SMC-deSolve} algorithms, we use $K = 500$ and $\phi = 0.999$ and utilize the same prior distributions and MCMC moves as those in the {\color{black} MCMC-deSolve methods}.   We run both MCMC algorithms with $400,000$ iterations, which is close to $K\cdot R$ in SMC-spline.

Figure \ref{fig:threecomparison} shows the comparison of four algorithms in terms of estimating $\thetabold$. Panel (a) and (b) show samples of the intermediate posterior distributions for $\thetabold$ by running SMC-spline and SMC-deSolve, respectively. {\color{black} The points with colors from light grey to dark grey are samples} for $r = 1, R/6, R/2, R$, where $R = 942$ in SMC-spline and $R = 1220$ in SMC-deSolve. 
 Panel (c) and (d) display the trace plots for $\thetabold$ by running MCMC-spline and MCMC-deSolve, respectively.  The black dots indicate the true parameter values in the ODE. For SMC-spline and SMC-deSolve, the particles gradually move to the posterior distribution with increasing annealing parameters. For MCMC-spline, the acceptance ratio of MH algorithm is $21.4\%$. It can explore one  mode rather than two modes  created in Equation (\ref{eq: ode2}). For MCMC-deSolve, the acceptance rate of MH algorithm is $25.2\%$. It gets stuck in local modes close to the initial value, and cannot explore the two modes. Table \ref{tab:comparison4methods} shows the posterior means of $\thetabold$ and $\xbold(0)$ as well as the corresponding $95\%$ credible interval (CI) using SMC-spline, SMC-deSolve, MCMC-spline, and MCMC-deSolve. 
SMC-spline and MCMC-spline provide posterior means close to the true values, and the corresponding credible intervals cover the true values, while MCMC-spline can only explore one mode of $\theta_1$. Our experiment in {\it Supplementary Section 4.1.1} reports the estimated ODE trajectories and the 95\% {\color{black} pointwise} credible bands by SMC-spline. The estimated mean ODE trajectories are very close to the true ODE trajectories. The 95\% {\color{black} pointwise} credible bands cover the true ODE trajectories.
The posterior mean of $\thetabold$ and $\xbold(0)$ provided by SMC-deSolve has a larger bias. The true value of  $\thetabold$ and $\xbold(0)$ are not included in the 95\% CIs, indicating that the estimated CI might be too narrow. MCMC-deSolve does not converge to the posterior distribution. 

\begin{figure}
\centering
\includegraphics[scale=0.85]{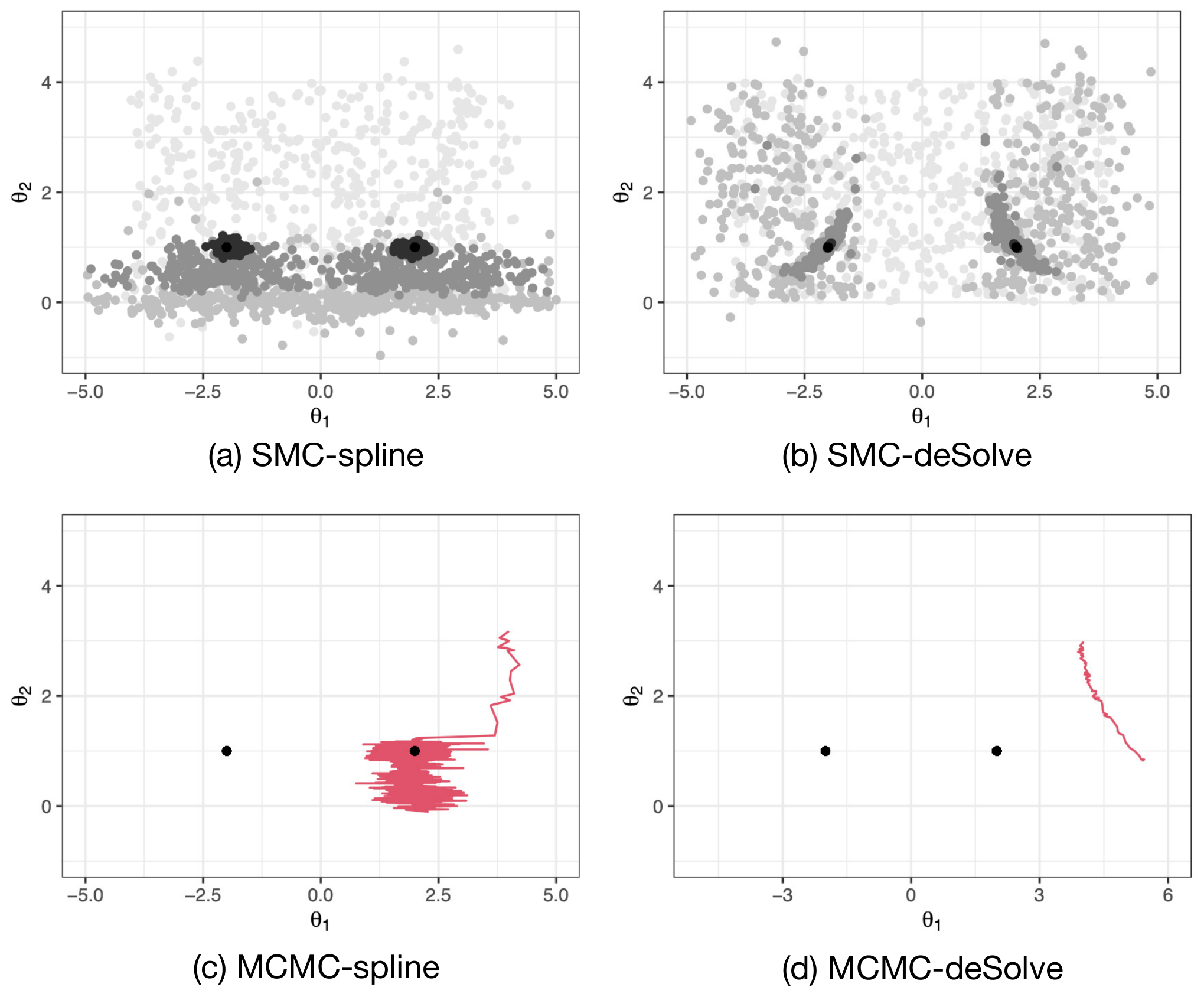}
\caption{(a) and (b): Intermediate posterior distributions for $\thetabold$ by running SMC-spline and SMC-deSolve, respectively. {\color{black} The points with colors from light grey to dark grey are samples for  $r = 1, R/6, R/2, R$.} (c) and (d): Trace plots for $\thetabold$ by running MCMC-spline and MCMC-deSolve, respectively. The black dots indicate true parameter values for generating ODEs. }
\label{fig:threecomparison}
\end{figure}

\begin{table}[ht]
\centering
\caption{Posterior mean of $\thetabold$ and $\xbold(0)$ ($95\%$ CI) from the four algorithms.}
\vspace{1em}
\label{tab:comparison4methods}
\begin{tabular}{cccccc}
  \hline
& True   &SMC-spline& SMC-deSolve & MCMC-spline & MCMC-deSolve \\ 
  \hline
$|\theta_{1}|$ & 2&1.93 (1.68, 2.19) & 1.84 (1.81, 1.88)& 1.92 (1.48, 2.37)  &5.37 (5.37, 5.37)\\ 
 $\theta_{2}$& 1& 0.99 (0.90, 1.09) & 1.12 (1.08, 1.17)& 0.97 (0.87, 1.08) &0.85 (0.85, 0.85)\\ 
 $x_{1}(0)$ & 7&6.43 (2.56, 10.11) & 3.94 (3.71, 4.22)& 6.48 (4.01, 8.92)  &4.55 (4.55, 4.55)\\ 
 $x_{2}(0)$& -10& -10.66 (-17.86, -2.26) & -4.47 (-5.63, -3.23)& -10.28 (-14.18,  -6.71) &0.68 (0.65, 0.70)\\ 
   \hline
\end{tabular}
\end{table}

We also run SMC-deSolve and MCMC-deSolve using the ``\emph{lsoda}" method in \emph{deSolve} with the same setting. Note that  the ``\emph{lsoda}" is the same ODE solver that is used for simulating ODEs and therefore it will favour the algorithms SMC-deSolve and MCMC-deSolve.   MCMC-deSolve does not converge to the posterior distribution. Table \ref{tab:newcomparison2methods} displays 
the posterior mean  of $\thetabold$ and $\xbold(0)$ together with the $95\%$ CI from SMC-deSolve and MCMC-deSolve. The posterior means from SMC-deSolve are close to the true values, and the credible intervals seem narrow. The true value of $\theta_2$ is on the boundary of the 95\% CI. Obviously, the methods with DE solvers heavily rely on the choice of  numerical solvers and they tend to ignore the uncertainty from approximating DE solutions using these solvers.

\begin{table}[ht]
\centering
\caption{Posterior mean of $\thetabold$ and $\xbold(0)$ together with the $95\%$ CI from SMC-deSolve and MCMC-deSolve using ``lsoda'' for solving ODE.}
\vspace{1em}
\label{tab:newcomparison2methods}
\begin{tabular}{ccccc}
  \hline
& True value  & SMC-deSolve &  MCMC-deSolve \\ 
  \hline
$|\theta_{1}|$ &2 & 1.98 (1.96, 2.01)  &1.56 (1.48, 1.65)\\ 
 $\theta_{2}$& 1 & 1.02 (1.00, 1.05) &4.58 (3.70, 7.03)\\ 
 $x_{1}(0)$ & 7 & 6.87 (6.61, 7.13)  &-0.09 (-0.51, 0.57)\\ 
 $x_{2}(0)$&  -10 & -10.52 (-11.58, -9.34) &6.65 (1.32, 12.45)\\ 
   \hline
\end{tabular}
\end{table}

We also compare two types of MCMC moves (MCMC-spline and MCMC-deSolve) in terms of computing time. Every $1000$ MCMC-spline moves cost $0.71$ seconds on a 2.3 GHz Intel Core i9 processor, while every $1000$ MCMC-deSolve moves cost $10.44$ seconds on same machine. This indicates that representing DE trajectories using a linear combination of basis functions can significantly increase the speed compared to using numerical solvers.

\subsubsection{Comparison of SMCs using different tuning parameters }
\label{sec:tuningparameters}
We conduct experiments investigating the selection of tuning parameters $\phi$, $K$, $\lambda$, and number of basis functions. 
{\it 1.} The parameter estimates get closer to the true values, and {\color{black} the RMSE} of the ODE trajectories gets smaller, when we increase the rCESS threshold $\phi$. A higher value of rCESS threshold is equivalent to more intermediate target distributions. {\it 2.} The proposed
SMC method performs better when we use a large number of particles. {\it 3.} For a given amount of computation, a relatively small $K$ and a large $\phi$ is optimal. However, a too small value of $K$ is not recommended, as an extremely small $K$ may lead to large Monte Carlo variance.
{\it 4.} This experiment indicates a sufficient number of basis functions is important in ODE trajectory estimation. However, we do not recommend using an overly large number of basis functions {\color{black} because it gains little in improving the approximation of ODE  trajectories and causes challenges in sampling high-dimensional basis function coefficients. }
{\it 5.} The Bayesian scheme for sampling $\lambda$ performs satisfactorily in terms of parameter estimates and estimated ODE trajectories.
The details of experiments are displayed in {\it Supplementary Section 4.1.2 and 4.1.3}.

\subsection{Delay differential equation examples}

\subsubsection{Hutchinson's equation}

Our first DDE example is the Hutchinson's equation, which is used to model the  blowfly data in \emph{Section} \ref{sec:real},
\begin{eqnarray*}
\frac{dx(t)}{dt} & = & \nu x(t)[1 - x(t-\tau)/(1000\cdot P)],
\end{eqnarray*}
where $\tau$, $\nu$, and $P$ are parameters of interest in the DDE. We set $x(0) = 3500$, $\tau = 3$, $\nu = 0.8$, and $P = 2$  to simulate the DDE trajectory. The DDE trajectory is observed with measurement error. The error is  lognormally distributed with mean $0$ and standard deviation $\sigma$.  We investigate the influence of the number of points per time step, and the influence of standard deviation of error $\sigma$. We simulate data sets in two scenarios. In the first scenario, we simulate $3$ data sets, with $101$, $201$, and $401$ observations respectively, equally spaced in $[0, 100]$. The standard deviation of error is $\sigma = 0.4$. In the second scenario, we simulate $36$ data sets with $201$ observations equally spaced in $[0, 100]$. The standard deviations of error for the $36$ data sets are $\sigma = (0.1, 0,5, 1.5)$, $12$ data sets for each level of $\sigma$.

We transform the positive constraint function $x(t) =e^{W(t)}$ and use B-spline basis functions to model the unconstrained function {\color{black} $W(t) = \Phibold(t)' \cbold$}. This is equivalent to solving the delay differential equation displayed in Equation (\ref{eq:12}).
We put $51$ knots equally spaced in $[0, 100]$, including $49$ interior knots. The total number of cubic B-spline functions is $L = 53$. 
The hyper-parameters in DDE parameter  prior/reference distributions and sequential Monte Carlo setups are the same as \emph{Section} \ref{sec:real}.

Table \ref{tab:3} displays the estimated parameters $(\nu, P, \tau, W(0))$ and RMSE defined in Equation (12) in {\it Supplementary} for $W(t)$ using data sets simulated in the first scenario. For the same DDE function, a larger number of observations improves the performance of estimation. Figure \ref{fig:DDE_comparison} shows the estimated parameters $(\nu, P, \tau, W(0))$ and RMSE using data sets simulated in the second scenario. It indicates that a smaller value of $\sigma$ improves the estimation.

\begin{table}[ht]
{
\begin{center}
\caption{Estimated parameters and MSE of $W(t)$ for three simulated data sets.}
\label{tab:3}

\begin{tabular}{c|c|c|c}
  \hline
 J &\centering $\nu$& P & $\tau$ \\ 
  \hline
   True& 0.8 & 2 & 3 \\ 
    101& 0.64 (0.43, 0.82) & 2.07 (1.56, 2.61)& 3.05 (2.66, 3.56) \\ 
 201& 0.73 (0.60, 0.90) & 1.98 (1.66, 2.32)& 3.02 (2.73, 3.26) \\ 
401& 0.75 (0.63, 0.86) & 2.09 (1.80, 2.38)&  3.00 (2.84, 3.16) \\ 
   \hline
  J &\centering  $W(0)$ &$\sigma^2$& RMSE\\ 
  \hline
   True& 8.16 & 0.16& $-$ \\ 
    101&  7.93 (7.34, 8.47) & 0.12 (0.08, 0.18) &0.26 \\ 
 201&  7.90 (7.53, 8.33)& 0.14 (0.11, 0.18) &0.21 \\ 
401&  7.98 (7.57, 8.40)& 0.17 (0.15, 0.19) &0.14 \\   
   \hline  
\end{tabular}
\end{center}
}
\end{table}

\begin{figure}[ht]
\center
\includegraphics[scale=0.6]{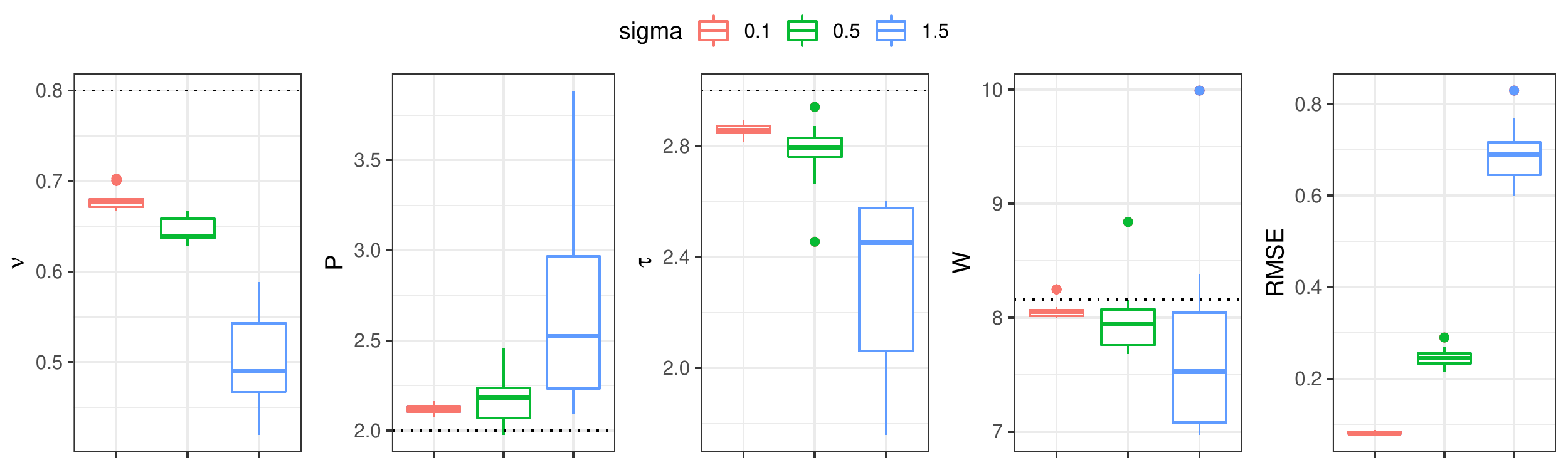}
\caption{Influence of $\sigma$ on DDE estimation. We simulate data sets with $\sigma = (0.1, 0.5, 1.5)$. A small value of $\sigma$ improves the performance of estimation. }
\label{fig:DDE_comparison}
\end{figure}

We also evaluate the quality of the uncertainty estimation of the parameters through a simulation study using Hutchinson's DDE model.  The true parameters are set to $\nu =0.22$, $P=2$, $\tau=8$,  and $\sigma = 0.2$. 
We generate 50 data sets of 200 observations in the time interval [0, 130] with different random seeds and run our {\color{black} SMC} algorithm for each of them.  We use 16 equally spaced knots in [0, 130].   
Table \ref{tab:CPC} shows the percentage that the 95\% CI's cover the true parameter value and the averaged 95\% CI. The coverage probability is close to the nominal 95\%.

\begin{table}[ht]
{
\begin{center}
\caption{Coverage probability and averaged value of the 95\% CI.}
\label{tab:CPC}

\begin{tabular}{|c|c|c|c|c|c|}
  \hline
 $K$&$\phi$ &\centering $\nu~(0.22) $& P~(2) & $\tau~(8)$ &  $\sigma$~(0.2) 
 \\ 
  \hline
 200 & 0.98 & 97.8\% & 97.8\%   & 91.1\%  & 95.5\%  \\  
  &  &  (0.20, 0.23) &  (1.91, 2.18)  &   (7.18, 8.18) &  (0.18, 0.22) 
 \\ \hline
\end{tabular}
\end{center}
}
\end{table}

\subsubsection{A nonlinear delay differential equation example}
In this section, we investigate a nonlinear delay differential equation model proposed by \cite{monk2003oscillatory} to model the feedback inhibition of gene expression. The nonlinear DDE is described as follows:
\begin{eqnarray}
\label{eq: ndde}
\frac{dx_{1}(t)}{dt} & = & \frac{1}{1+(x_{2}(t-\tau)/p_{0})^{n}}-\mu_{m}x_{1}(t), \nonumber\\
\frac{dx_{2}(t)}{dt} & = & x_{1}(t)-\mu_{p}x_{2}(t).
\end{eqnarray}
In Equation (\ref{eq: ndde}), $x_{1}(t)$ denotes the expression of \emph{mRNA} at time $t$, and $x_{2}(t)$ denotes the expression of a \emph{protein} at time $t$. There is a delayed repression of \emph{mRNA} production by the \emph{protein}. The DDE system depends on the \emph{transcriptional delay} $\tau$, and degradation rates $\mu_{m}$ and $\mu_{p}$, the expression threshold $p_{0}$, and the Hill coefficient $n$. As noted in \cite{monk2003oscillatory}, there is significant nonlinearity in the DDE system when the Hill coefficient $n > 4$. 
We simulate a delay differential equation system and noisy observations. We use B-spline functions to represent the DDE trajectories and use SMC to estimate parameters. The {\color{black} posterior} means of the parameters are fairly close to the true values, and the 95\% credible intervals cover the true values. The estimated mean DDE trajectories are generally very close to the true DDE trajectories. The 95\% {\color{black} pointwise} confidence bands cover the true DDE trajectories. The details of setups and results are shown in {\it Supplementary Section 4.2}.

\section{Discussion}
\label{sec:conc}

We proposed an adaptive semi-parametric Bayesian framework to solve nonlinear differential equations and estimate the DE parameters using an efficient annealed sequential Monte Carlo method. The main idea is to represent DE trajectories using a linear combination of basis functions and to estimate the coefficients of these basis functions together with other DE parameters using an annealed sequential Monte Carlo algorithm. The proposed method avoids using  DE solvers which can be computationally expensive and sensitive to the initial state and model parameters. Our work is a Bayesian method with two obvious advantages over the counterpart using a {\color{black} frequentist} method. First, the Bayesian method can easily achieve uncertainty estimates. Second, we avoid expensive tuning for the global smoothing parameter by treating it in the same way as other parameters.

We represent DE variables with a linear combination of basis functions. 
A prior distribution on the basis function coefficients
is used to control the trade-off between fit to the data and fidelity to the DE model. 
Our model is related to the generalized profiling approaches developed by \cite{ramsay2007parameter}, in which the coefficients of the basis functions and DE parameters are estimated by a penalized smoothing procedure. \cite{qi2010asymptotic} investigated the asymptotic bias induced by the spline approximation in the generalized profiling approaches. \cite{pang2017asymptotically} loosened the assumptions for the asymptotic properties. 
We refer readers to \cite{qi2010asymptotic, pang2017asymptotically} for the details of assumptions, theorems, and proofs.

We developed a sequential Monte Carlo method in an annealing framework to estimate the DE parameters.  The annealed SMC considers the same parameter space for all the intermediate distributions.
Consequently,  MCMC moves used in the literature on Bayesian inference for differential equations can be repurposed to act as SMC proposal distributions in the annealed SMC. Note that more advanced MCMC moves can be used to further improve the performance of the annealed SMC.  Annealed SMC is preferred over MCMC algorithms for several reasons. First, the developed SMC method can fully explore the multi-modal posterior surface of {\color{black} DE} parameters. 
Second, the proposed method is a semi-automatic algorithm that requires minimal tuning from the user; given a criterion for the relative conditional effective sample size and the number of particles, it can adaptively choose a scheme for the sequence of the annealing parameters that determine the intermediate target distributions of SMC.   
Third, the annealed SMC is an embarrassingly parallel method. Unlike running an MCMC chain for a long time until it converges, the annealed SMC is more efficient because a large number of particles can be run on different CPUs or GPUs simultaneously.

We used different simulation scenarios to explore the numerical behavior of our model and method, and demonstrated it can perform well in both ODE and DDE parameter estimation. Our simulation studies provide some guideline for choosing the value of rCESS and the number of particles. To ensure more accurate estimates of the DE parameters from the annealed SMC, a rule of thumb is to choose a large number for rCESS and to avoid using an extremely small number of particles.  We also applied our method to a real data example to model the population dynamics of blowflies with a delay differential equation. The delay parameter in DDEs is usually challenging to estimate. But our application shows that our method is superior to the previous frequentist method.

There are several improvements and extensions based on our proposed method for future work. 
\textcolor{black}{ Note there might be other satisfactory ways to construct the intermediate distributions for our SMC algorithm such that  the last distribution of the sequence is our target distribution.  In theory,  when the number of particles and the number of the SMC iterations are large enough, such an alternative SMC algorithm can also  well  approximate the same target distribution. But in practice,  different sequences of intermediate distributions  may perform differently  given the same computing budget. In future, it is worth exploring alternative ways to construct the intermediate distributions and make comparisons. }
For simplicity, we have used the same reference distributions as the prior distributions for most of the parameters. The performance of the annealed SMC can be improved by using  reference distributions that are close to the target distribution.   
In all of our current numerical experiments, we put equally spaced knots for smoothing splines and the number of knots are pre-determined before running experiments. In future work, we will explore using a smaller number of knots that are well placed, and let the data determine the number of knots and their locations. The adaptive control of knots in smoothing spline for DEs will benefit the estimation of DEs, especially those with sharp changes. 
In practice, it is often the case that there are several DE models that are proposed to describe the same dynamic system. This requires selection among various differential equations models. One direction of future work is to explore model selection methods  for DEs. Another line of future work is to develop more scalable SMC algorithms for estimating parameters in a large series of differential equations.

\section{Supplementary Materials}

\begin{description}
\item[Technical material:] The file supplementary.pdf (PDF file) provides details of the algorithms and simulation studies in the article. 
\item[R-package:]  The R package ``smcDE''  was developed  to implement our proposed methods, real data analysis, and simulation studies. It is available from \url{https://github.com/shijiaw/smcDE}. 
\end{description}



\begin{thebibliography}{}

\bibitem[\protect\citeauthoryear{Ascher, Ruuth, and Spiteri}{Ascher
  et~al.}{1997}]{ascher1997implicit}
Ascher, U.~M., S.~J. Ruuth, and R.~J. Spiteri (1997).
\newblock Implicit-explicit {R}unge-{K}utta methods for time-dependent partial
  differential equations.
\newblock {\em Applied Numerical Mathematics\/}~{\em 25\/}(2-3), 151--167.

\bibitem[\protect\citeauthoryear{Barber and Wang}{Barber and
  Wang}{2014}]{barber2014gaussian}
Barber, D. and Y.~Wang (2014).
\newblock Gaussian processes for {B}ayesian estimation in ordinary differential
  equations.
\newblock In {\em International Conference on Machine Learning}, pp.\
  1485--1493.

\bibitem[\protect\citeauthoryear{Berezansky, Braverman, and Idels}{Berezansky
  et~al.}{2010}]{berezansky2010nicholson}
Berezansky, L., E.~Braverman, and L.~Idels (2010).
\newblock Nicholson's blowflies differential equations revisited: main results
  and open problems.
\newblock {\em Applied Mathematical Modelling\/}~{\em 34\/}(6), 1405--1417.

\bibitem[\protect\citeauthoryear{Berry, Carroll, and Ruppert}{Berry
  et~al.}{2002}]{berry2002bayesian}
Berry, S.~M., R.~J. Carroll, and D.~Ruppert (2002).
\newblock Bayesian smoothing and regression splines for measurement error
  problems.
\newblock {\em Journal of the American Statistical Association\/}~{\em
  97\/}(457), 160--169.

\bibitem[\protect\citeauthoryear{Bhaumik, Ghosal, et~al.}{Bhaumik
  et~al.}{2015}]{bhaumik2015bayesian}
Bhaumik, P., S.~Ghosal, et~al. (2015).
\newblock Bayesian two-step estimation in differential equation models.
\newblock {\em Electronic Journal of Statistics\/}~{\em 9\/}(2), 3124--3154.

\bibitem[\protect\citeauthoryear{Bhaumik, Ghosal, et~al.}{Bhaumik
  et~al.}{2017}]{bhaumik2017efficient}
Bhaumik, P., S.~Ghosal, et~al. (2017).
\newblock Efficient {B}ayesian estimation and uncertainty quantification in
  ordinary differential equation models.
\newblock {\em Bernoulli\/}~{\em 23\/}(4B), 3537--3570.

\bibitem[\protect\citeauthoryear{Bulirsch and Stoer}{Bulirsch and
  Stoer}{1966}]{bulirsch1966numerical}
Bulirsch, R. and J.~Stoer (1966).
\newblock Numerical treatment of ordinary differential equations by
  extrapolation methods.
\newblock {\em Numerische Mathematik\/}~{\em 8\/}(1), 1--13.

\bibitem[\protect\citeauthoryear{Burden, Faires, and Reynolds}{Burden
  et~al.}{2001}]{burden2001numerical}
Burden, R.~L., J.~D. Faires, and A.~C. Reynolds (2001).
\newblock Numerical analysis.

\bibitem[\protect\citeauthoryear{Butcher}{Butcher}{2016}]{butcher2016numerical}
Butcher, J.~C. (2016).
\newblock {\em Numerical methods for ordinary differential equations}.
\newblock John Wiley \& Sons.

\bibitem[\protect\citeauthoryear{Calderhead, Girolami, and Lawrence}{Calderhead
  et~al.}{2009}]{calderhead2009accelerating}
Calderhead, B., M.~Girolami, and N.~D. Lawrence (2009).
\newblock Accelerating {B}ayesian inference over nonlinear differential
  equations with {G}aussian processes.
\newblock In {\em Advances in Neural Information Processing Systems}, pp.\
  217--224.

\bibitem[\protect\citeauthoryear{Campbell and Steele}{Campbell and
  Steele}{2012}]{campbell2012smooth}
Campbell, D. and R.~J. Steele (2012).
\newblock Smooth functional tempering for nonlinear differential equation
  models.
\newblock {\em Statistics and Computing\/}~{\em 22\/}(2), 429--443.

\bibitem[\protect\citeauthoryear{Cao, Huang, and Wu}{Cao
  et~al.}{2012}]{cao2012penalized}
Cao, J., J.~Z. Huang, and H.~Wu (2012).
\newblock Penalized nonlinear least squares estimation of time-varying
  parameters in ordinary differential equations.
\newblock {\em Journal of Computational and Graphical Statistics\/}~{\em
  21\/}(1), 42--56.

\bibitem[\protect\citeauthoryear{Cao, Wang, and Xu}{Cao
  et~al.}{2011}]{cao2011robust}
Cao, J., L.~Wang, and J.~Xu (2011).
\newblock Robust estimation for ordinary differential equation models.
\newblock {\em Biometrics\/}~{\em 67\/}(4), 1305--1313.

\bibitem[\protect\citeauthoryear{Carpenter, Clifford, and Fearnhead}{Carpenter
  et~al.}{1999}]{carpenter1999improved}
Carpenter, J., P.~Clifford, and P.~Fearnhead (1999).
\newblock Improved particle filter for nonlinear problems.
\newblock {\em IEE Proceedings-Radar, Sonar and Navigation\/}~{\em 146\/}(1),
  2--7.

\bibitem[\protect\citeauthoryear{Chen and Wu}{Chen and
  Wu}{2008}]{chen2008efficient}
Chen, J. and H.~Wu (2008).
\newblock Efficient local estimation for time-varying coefficients in
  deterministic dynamic models with applications to {HIV}-1 dynamics.
\newblock {\em Journal of the American Statistical Association\/}~{\em
  103\/}(481), 369--384.

\bibitem[\protect\citeauthoryear{Chopin et~al.}{Chopin
  et~al.}{2004}]{chopin2004central}
Chopin, N. et~al. (2004).
\newblock Central limit theorem for sequential {M}onte {C}arlo methods and its
  application to {B}ayesian inference.
\newblock {\em The Annals of Statistics\/}~{\em 32\/}(6), 2385--2411.

\bibitem[\protect\citeauthoryear{Dass, Lee, Lee, and Park}{Dass
  et~al.}{2017}]{dass2017laplace}
Dass, S.~C., J.~Lee, K.~Lee, and J.~Park (2017).
\newblock Laplace based approximate posterior inference for differential
  equation models.
\newblock {\em Statistics and Computing\/}~{\em 27\/}(3), 679--698.

\bibitem[\protect\citeauthoryear{De~Boor}{De~Boor}{1972}]{de1972calculating}
De~Boor, C. (1972).
\newblock On calculating with {B}-splines.
\newblock {\em Journal of Approximation Theory\/}~{\em 6\/}(1), 50--62.

\bibitem[\protect\citeauthoryear{Del~Moral}{Del~Moral}{2004}]{del2004feynman}
Del~Moral, P. (2004).
\newblock {\em Feynman-Kac Formulae: Genealogical and Interacting Particle
  Systems with Applications}.
\newblock New York: Springer.

\bibitem[\protect\citeauthoryear{Del~Moral, Doucet, and Jasra}{Del~Moral
  et~al.}{2006}]{del2006sequential}
Del~Moral, P., A.~Doucet, and A.~Jasra (2006).
\newblock Sequential {M}onte {C}arlo samplers.
\newblock {\em Journal of the Royal Statistical Society: Series B (Statistical
  Methodology)\/}~{\em 68\/}(3), 411--436.

\bibitem[\protect\citeauthoryear{Del~Moral, Doucet, and Jasra}{Del~Moral
  et~al.}{2012}]{del2012adaptive}
Del~Moral, P., A.~Doucet, and A.~Jasra (2012).
\newblock An adaptive sequential {M}onte {C}arlo method for approximate
  {B}ayesian computation.
\newblock {\em Statistics and Computing\/}~{\em 22\/}(5), 1009--1020.

\bibitem[\protect\citeauthoryear{Dondelinger, Husmeier, Rogers, and
  Filippone}{Dondelinger et~al.}{2013}]{dondelinger2013ode}
Dondelinger, F., D.~Husmeier, S.~Rogers, and M.~Filippone (2013).
\newblock {ODE} parameter inference using adaptive gradient matching with
  {G}aussian processes.
\newblock In {\em Artificial Intelligence and Statistics}, pp.\  216--228.

\bibitem[\protect\citeauthoryear{Douc and Capp{\'e}}{Douc and
  Capp{\'e}}{2005}]{douc2005comparison}
Douc, R. and O.~Capp{\'e} (2005).
\newblock Comparison of resampling schemes for particle filtering.
\newblock In {\em Image and Signal Processing and Analysis, 2005. ISPA 2005.
  Proceedings of the 4th International Symposium on}, pp.\  64--69. IEEE.

\bibitem[\protect\citeauthoryear{Doucet, De~Freitas, and Gordon}{Doucet
  et~al.}{2001}]{doucet2001introduction}
Doucet, A., N.~De~Freitas, and N.~Gordon (2001).
\newblock An introduction to sequential {M}onte {C}arlo methods.
\newblock In {\em Sequential Monte Carlo methods in practice}, pp.\  3--14.
  Springer.

\bibitem[\protect\citeauthoryear{Doucet, Godsill, and Andrieu}{Doucet
  et~al.}{2000}]{doucet2000sequential}
Doucet, A., S.~Godsill, and C.~Andrieu (2000).
\newblock On sequential {M}onte {C}arlo sampling methods for {B}ayesian
  filtering.
\newblock {\em Statistics and Computing\/}~{\em 10\/}(3), 197--208.

\bibitem[\protect\citeauthoryear{Fan, Wu, Chen, Kuo, and Lewis}{Fan
  et~al.}{2011}]{fan2011choosing}
Fan, Y., R.~Wu, M.-H. Chen, L.~Kuo, and P.~O. Lewis (2011).
\newblock Choosing among partition models in {B}ayesian phylogenetics.
\newblock {\em Molecular Biology and Evolution\/}~{\em 28\/}(1), 523--532.

\bibitem[\protect\citeauthoryear{Hochbruck, Lubich, and Selhofer}{Hochbruck
  et~al.}{1998}]{hochbruck1998exponential}
Hochbruck, M., C.~Lubich, and H.~Selhofer (1998).
\newblock Exponential integrators for large systems of differential equations.
\newblock {\em SIAM Journal on Scientific Computing\/}~{\em 19\/}(5),
  1552--1574.

\bibitem[\protect\citeauthoryear{Hochbruck and Ostermann}{Hochbruck and
  Ostermann}{2010}]{hochbruck2010exponential}
Hochbruck, M. and A.~Ostermann (2010).
\newblock Exponential integrators.
\newblock {\em Acta Numerica\/}~{\em 19}, 209--286.

\bibitem[\protect\citeauthoryear{Hol, Schon, and Gustafsson}{Hol
  et~al.}{2006}]{hol2006resampling}
Hol, J.~D., T.~B. Schon, and F.~Gustafsson (2006).
\newblock On resampling algorithms for particle filters.
\newblock In {\em Nonlinear Statistical Signal Processing Workshop, 2006 IEEE},
  pp.\  79--82. IEEE.

\bibitem[\protect\citeauthoryear{Jain}{Jain}{1979}]{jain1979numerical}
Jain, M.~K. (1979).
\newblock {\em Numerical solution of differential equations}.
\newblock Wiley Eastern New Delhi.

\bibitem[\protect\citeauthoryear{Jameson, Schmidt, and Turkel}{Jameson
  et~al.}{1981}]{jameson1981numerical}
Jameson, A., W.~Schmidt, and E.~Turkel (1981).
\newblock Numerical solution of the {E}uler equations by finite volume methods
  using {R}unge {K}utta time stepping schemes.
\newblock In {\em 14th fluid and plasma dynamics conference}, pp.\  1259.

\bibitem[\protect\citeauthoryear{Kitagawa}{Kitagawa}{1996}]{kitagawa1996monte}
Kitagawa, G. (1996).
\newblock Monte {C}arlo filter and smoother for non-{G}aussian nonlinear state
  space models.
\newblock {\em Journal of Computational and Graphical Statistics\/}~{\em
  5\/}(1), 1--25.

\bibitem[\protect\citeauthoryear{Lee, Lee, and Dass}{Lee
  et~al.}{2018}]{lee2018inference}
Lee, K., J.~Lee, and S.~C. Dass (2018).
\newblock Inference for differential equation models using relaxation via
  dynamical systems.
\newblock {\em Computational Statistics \& Data Analysis\/}~{\em 127},
  116--134.

\bibitem[\protect\citeauthoryear{Liu and Chen}{Liu and
  Chen}{1998}]{liu1998sequential}
Liu, J.~S. and R.~Chen (1998).
\newblock Sequential {M}onte {C}arlo methods for dynamic systems.
\newblock {\em Journal of the American Statistical Association\/}~{\em
  93\/}(443), 1032--1044.

\bibitem[\protect\citeauthoryear{May}{May}{1976}]{may1976models}
May, R.~M. (1976).
\newblock Models for single populations.
\newblock {\em Theoretical Ecology\/}.

\bibitem[\protect\citeauthoryear{Monk}{Monk}{2003}]{monk2003oscillatory}
Monk, N.~A. (2003).
\newblock Oscillatory expression of {H}es1, p53, and {NF}-$\kappa${B} driven by
  transcriptional time delays.
\newblock {\em Current Biology\/}~{\em 13\/}(16), 1409--1413.

\bibitem[\protect\citeauthoryear{Neal}{Neal}{2001}]{neal2001annealed}
Neal, R.~M. (2001).
\newblock Annealed importance sampling.
\newblock {\em Statistics and Computing\/}~{\em 11\/}(2), 125--139.

\bibitem[\protect\citeauthoryear{Nicholson}{Nicholson}{1954}]{nicholson1954outline}
Nicholson, A.~J. (1954).
\newblock An outline of the dynamics of animal populations.
\newblock {\em Australian Journal of Zoology\/}~{\em 2\/}(1), 9--65.

\bibitem[\protect\citeauthoryear{Pang, Yan, and Zhou}{Pang
  et~al.}{2017}]{pang2017asymptotically}
Pang, T., P.~Yan, and H.~H. Zhou (2017).
\newblock Asymptotically efficient parameter estimation for ordinary
  differential equations.
\newblock {\em Science China Mathematics\/}~{\em 60\/}(11), 2263--2286.

\bibitem[\protect\citeauthoryear{Poyton, Varziri, McAuley, McLellan, and
  Ramsay}{Poyton et~al.}{2006}]{poyton2006parameter}
Poyton, A., M.~S. Varziri, K.~B. McAuley, P.~McLellan, and J.~O. Ramsay (2006).
\newblock Parameter estimation in continuous-time dynamic models using
  principal differential analysis.
\newblock {\em Computers \& Chemical Engineering\/}~{\em 30\/}(4), 698--708.

\bibitem[\protect\citeauthoryear{Qi, Zhao, et~al.}{Qi
  et~al.}{2010}]{qi2010asymptotic}
Qi, X., H.~Zhao, et~al. (2010).
\newblock Asymptotic efficiency and finite-sample properties of the generalized
  profiling estimation of parameters in ordinary differential equations.
\newblock {\em The Annals of Statistics\/}~{\em 38\/}(1), 435--481.

\bibitem[\protect\citeauthoryear{Ramsay}{Ramsay}{2004}]{ramsay2004functional}
Ramsay, J.~O. (2004).
\newblock Functional data analysis.
\newblock {\em Encyclopedia of Statistical Sciences\/}~{\em 4}.

\bibitem[\protect\citeauthoryear{Ramsay, Hooker, Campbell, and Cao}{Ramsay
  et~al.}{2007}]{ramsay2007parameter}
Ramsay, J.~O., G.~Hooker, D.~Campbell, and J.~Cao (2007).
\newblock Parameter estimation for differential equations: a generalized
  smoothing approach.
\newblock {\em Journal of the Royal Statistical Society: Series B (Statistical
  Methodology)\/}~{\em 69\/}(5), 741--796.

\bibitem[\protect\citeauthoryear{Ramsay and Silverman}{Ramsay and
  Silverman}{2007}]{ramsay2007applied}
Ramsay, J.~O. and B.~W. Silverman (2007).
\newblock {\em Applied functional data analysis: methods and case studies}.
\newblock Springer.

\bibitem[\protect\citeauthoryear{Reiss and Todd~Ogden}{Reiss and
  Todd~Ogden}{2009}]{reiss2009smoothing}
Reiss, P.~T. and R.~Todd~Ogden (2009).
\newblock Smoothing parameter selection for a class of semiparametric linear
  models.
\newblock {\em Journal of the Royal Statistical Society: Series B (Statistical
  Methodology)\/}~{\em 71\/}(2), 505--523.

\bibitem[\protect\citeauthoryear{Rosenzweig and MacArthur}{Rosenzweig and
  MacArthur}{1963}]{rosenzweig1963graphical}
Rosenzweig, M.~L. and R.~H. MacArthur (1963).
\newblock Graphical representation and stability conditions of predator-prey
  interactions.
\newblock {\em The American Naturalist\/}~{\em 97\/}(895), 209--223.

\bibitem[\protect\citeauthoryear{Soetaert, Petzoldt, and Setzer}{Soetaert
  et~al.}{2010}]{soetaert2010solving}
Soetaert, K., T.~Petzoldt, and R.~W. Setzer (2010).
\newblock Solving differential equations in {R}: package desolve.
\newblock {\em Journal of Statistical Software\/}~{\em 33}.

\bibitem[\protect\citeauthoryear{Varah}{Varah}{1982}]{varah1982spline}
Varah, J.~M. (1982).
\newblock A spline least squares method for numerical parameter estimation in
  differential equations.
\newblock {\em SIAM Journal on Scientific and Statistical Computing\/}~{\em
  3\/}(1), 28--46.

\bibitem[\protect\citeauthoryear{Wang and Cao}{Wang and
  Cao}{2012}]{wang2012estimating}
Wang, L. and J.~Cao (2012).
\newblock Estimating parameters in delay differential equation models.
\newblock {\em Journal of Agricultural, Biological, and Environmental
  Statistics\/}~{\em 17\/}(1), 68--83.

\bibitem[\protect\citeauthoryear{Wang, Wang, and Bouchard-C{\^o}t{\'e}}{Wang
  et~al.}{2020}]{wang2018annealed}
Wang, L., S.~Wang, and A.~Bouchard-C{\^o}t{\'e} (2020).
\newblock An annealed sequential {M}onte {C}arlo method for {B}ayesian
  phylogenetics.
\newblock {\em Systematic Biology\/}~{\em 69\/}(1), 155--183.

\bibitem[\protect\citeauthoryear{Wood}{Wood}{2017}]{wood2017generalized}
Wood, S.~N. (2017).
\newblock {\em Generalized additive models: an introduction with {R}}.
\newblock CRC press.

\bibitem[\protect\citeauthoryear{Zhang, Yin, Caffo, Sun, Boatman-Reich,
  et~al.}{Zhang et~al.}{2017}]{zhang2017bayesian}
Zhang, T., Q.~Yin, B.~Caffo, Y.~Sun, D.~Boatman-Reich, et~al. (2017).
\newblock Bayesian inference of high-dimensional, cluster-structured ordinary
  differential equation models with applications to brain connectivity studies.
\newblock {\em The Annals of Applied Statistics\/}~{\em 11\/}(2), 868--897.

\bibitem[\protect\citeauthoryear{Zhou, Johansen, and Aston}{Zhou
  et~al.}{2016}]{zhou2016toward}
Zhou, Y., A.~M. Johansen, and J.~A. Aston (2016).
\newblock Toward automatic model comparison: an adaptive sequential {M}onte
  {C}arlo approach.
\newblock {\em Journal of Computational and Graphical Statistics\/}~{\em
  25\/}(3), 701--726.

\end{thebibliography}


\begin{thebibliography}{}

\bibitem[\protect\citeauthoryear{Chopin et~al.}{Chopin
  et~al.}{2004}]{chopin2004central}
Chopin, N. et~al. (2004).
\newblock Central limit theorem for sequential {M}onte {C}arlo methods and its
  application to {B}ayesian inference.
\newblock {\em The Annals of Statistics\/}~{\em 32\/}(6), 2385--2411.

\bibitem[\protect\citeauthoryear{Del~Moral, Doucet, and Jasra}{Del~Moral
  et~al.}{2006}]{del2006sequential}
Del~Moral, P., A.~Doucet, and A.~Jasra (2006).
\newblock Sequential {M}onte {C}arlo samplers.
\newblock {\em Journal of the Royal Statistical Society: Series B (Statistical
  Methodology)\/}~{\em 68\/}(3), 411--436.

\bibitem[\protect\citeauthoryear{Monk}{Monk}{2003}]{monk2003oscillatory}
Monk, N.~A. (2003).
\newblock Oscillatory expression of {H}es1, p53, and {NF}-$\kappa${B} driven by
  transcriptional time delays.
\newblock {\em Current Biology\/}~{\em 13\/}(16), 1409--1413.

\bibitem[\protect\citeauthoryear{Soetaert, Petzoldt, and Setzer}{Soetaert
  et~al.}{2010}]{soetaert2010solving}
Soetaert, K., T.~Petzoldt, and R.~W. Setzer (2010).
\newblock Solving differential equations in {R}: package desolve.
\newblock {\em Journal of Statistical Software\/}~{\em 33}.

\bibitem[\protect\citeauthoryear{Vince and Vince}{Vince and
  Vince}{2010}]{vince2010mathematics}
Vince, J. and J.~A. Vince (2010).
\newblock {\em Mathematics for computer graphics}, Volume~3.
\newblock Springer.

\bibitem[\protect\citeauthoryear{Wang, Wang, and Bouchard-C{\^o}t{\'e}}{Wang
  et~al.}{2020}]{wang2018annealed}
Wang, L., S.~Wang, and A.~Bouchard-C{\^o}t{\'e} (2020).
\newblock An annealed sequential {M}onte {C}arlo method for {B}ayesian
  phylogenetics.
\newblock {\em Systematic Biology\/}~{\em 69\/}(1), 155--183.

\end{thebibliography}
\end{document}


\def\spacingset#1{\renewcommand{\baselinestretch}%
{#1}\small\normalsize} \spacingset{1}


\if0\blind
{
  \title{\bf Supplementary Material for ``Adaptive semiparametric Bayesian differential equations via sequential Monte Carlo''}
 \author{Shijia Wang, Shufei Ge, Renny Doig, Liangliang Wang}
  \maketitle
} \fi

\if1\blind
{
  \bigskip
  \bigskip
  \bigskip
  \begin{center}
    {\LARGE\bf Title}
\end{center}
  \medskip
} \fi

\bigskip

\spacingset{1.45}

\begin{figure}
\center
\includegraphics[scale=0.8]{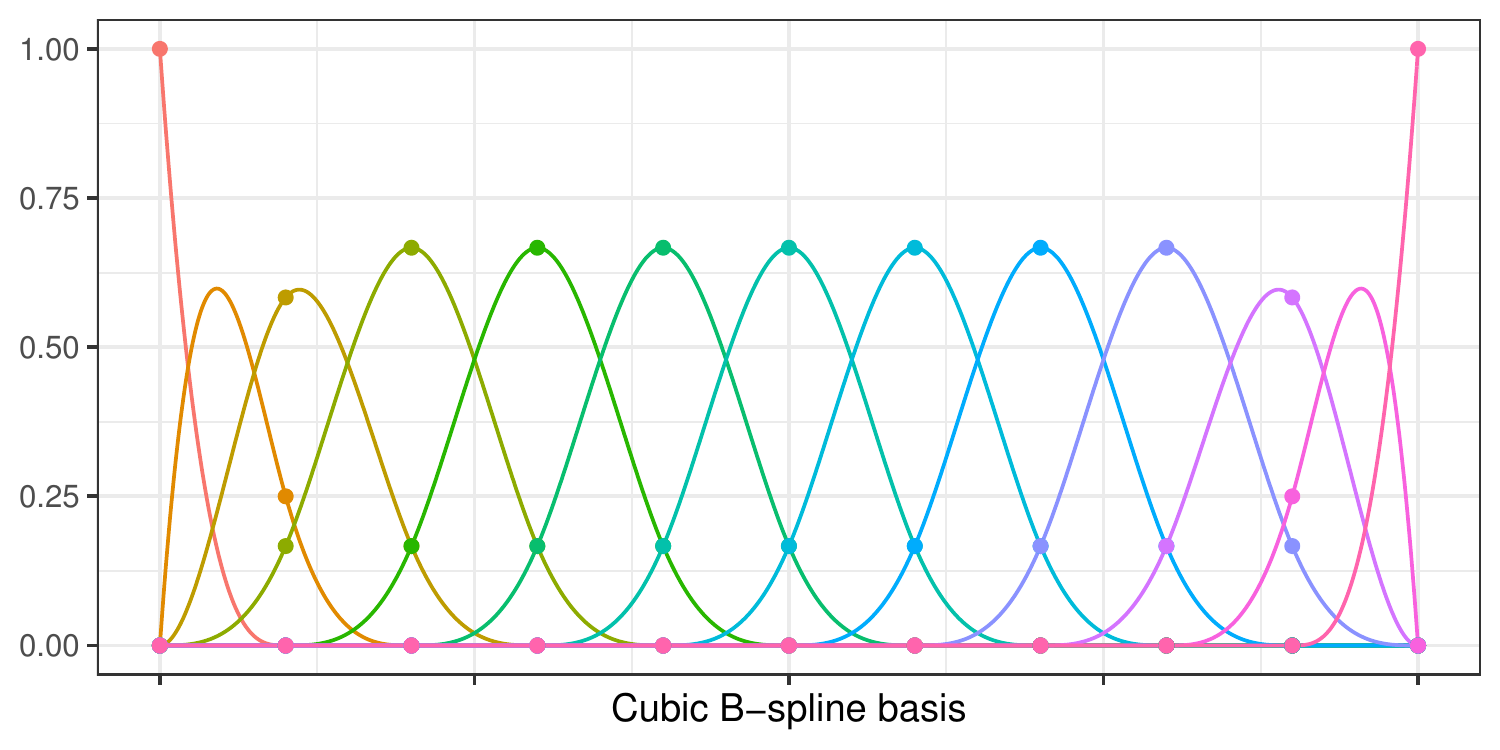}
\caption{ The thirteen B-spline basis functions defined on [0, 1] with degree $d= 3$ and $9$ 
equally spaced knots; each basis function
is positive over at most $d + 1$ adjacent subintervals. The continuity characteristics of cubic B-spline functions make the segment joints smooth \citep{vince2010mathematics}. } 
\label{fig:basisfunction}
\end{figure}

\section{ Propagation step of annealed SMC}
We propagate new samples $\{\betabold_{r}^{(k)}\}_{k = 1}^K$ via $\pi_{r}$-invariant MCMC moves, $\{\betabold_{r}^{(k)} \sim T_{r}(\tilde\betabold_{r-1}^{(k)}, \cdot)\}_{k = 1}^K$. The full conditional posterior distributions, $\pi_{r}(\sigma_{i}^{2}|\cbold_{i})$, $\pi_{r}( \tau |\cbold, \thetabold, \lambda)$, $\pi_{r}( \thetabold |\cbold,\tau, \lambda)${\color{black} , $\pi_{r}( \lambda |\cbold,\tau, \thetabold)$} and $\pi_{r}( \cbold_{i} |\tau, \thetabold, \boldsymbol{\sigma}, \cbold_{-i}, \lambda)$ admit the following forms:

\begin{itemize}
\item The full conditional distribution of $\sigma_{i}^{2}$ is 
\begin{eqnarray}
\label{eq:sigmar}
\sigma_{i}^{2}|\cbold_{i}\sim IG\bigg(g_{0}+\frac{{\color{black} J_i}}{2}, h_{0}+\frac{\alpha_{r}}{2}\sum_{j = 1}^{{\color{black} J_i}}(y_{ij} - \Phibold_i(t_{ij})' \cbold_{i})^{2}\bigg).
\end{eqnarray}

\item The full conditional distribution $\pi_{r}( \tau |\cbold, \thetabold, \lambda)$ does not admit a closed form. 
\begin{eqnarray}
\label{eq:taur}
&&\gamma_{r}( \tau |\cbold, \thetabold, \lambda) 
\propto \\ &&\exp\bigg\{ -\alpha_{r}\sum_{i=1}^{I}\bigg(\frac{\lambda}{2}\sum_{l_i = 0}^{\zeta_i}\sum_{m=1}^{M}v_{l_i m}\cdot\bigg(\bigg[\frac{d\Phibold_i(s)'}{ds} \cbold_{i}-g_{i}({\color{black} \Phibold(s)' \cbold , \Phibold(s-\tau)' \cbold}|\thetabold) \bigg]^{2}\bigg|_{s = \xi_{il_im}}\bigg)\bigg)
\bigg\}. \nonumber
\end{eqnarray}

We perform a random walk MH algorithm with a Gaussian kernel to  propose $\tau$. 
\begin{enumerate}
\item  $\tau^{\star}\sim N(\tau, \sigma_{\tau}^{2})\text{I}(\tau^{\star}>0)$, 
\item compute the acceptance probability 
\[
p_{\mathit{MH}} = \min\bigg\{1, \frac{\gamma_{r}(\tau^{\star}|\cbold, \thetabold, \lambda)}{\gamma_{r}(\tau|\cbold, \thetabold, \lambda)}\bigg\},
\]
\item sample $u\sim U(0,1)$, we accept $\tau = \tau^{\star}$ if $u<p_{MH}$, otherwise we set $\tau = \tau^*$.
\end{enumerate}

\item 
The existence of closed-form conditional posterior distributions  $\pi_{r}(\theta_{d}|\cbold, \tau, \lambda)$ and $\pi_{r}(\cbold_{i} |\thetabold, \boldsymbol{\sigma},  \cbold_{-i}, \lambda)$ depends on $g_{i}$ $(i = 1, 2, \ldots, I)$. If all $g_{i}$ are linear functions of $\theta_{d}$ (or $\cbold_{i}$), there exists closed-form conditional posterior distribution $\pi_{r}(\theta_{d}|\cbold, \tau, \lambda)$ (or $\pi_{r}(\cbold_{i} |\thetabold, \boldsymbol{\sigma},  \cbold_{-i}, \lambda)$), which is Gaussian distributed. Otherwise, we perform a random walk MH algorithm with a Gaussian kernel. 

\item If the full conditional distribution $\pi_{r}(\cbold_{i} |\tau, \thetabold, \boldsymbol{\sigma}, \cbold_{-i}, \lambda)$ does not admit a closed form. We perform a random walk MH algorithm with a Gaussian kernel with
{\small
\begin{eqnarray}
\label{eq:cr}
&&\gamma_{r}( \cbold_{i} |\tau, \thetabold, \boldsymbol{\sigma}, \cbold_{-i}, \lambda)\propto \nonumber\\
 &&\exp\bigg\{ -\alpha_{r} \bigg( \textcolor{black}{\sum_{i  \in \mathcal{I}_0} }\sum_{j=1}^{{\color{black} J_i}}\frac{(y_{ij} -  \Phibold_i(t_{ij})' \cbold_{i})^{2}}{2\sigma_{i}^{2}} +\nonumber\\
&&  \frac{\lambda}{2} \textcolor{black}{\sum_{i=1}^{I}} \sum_{l_i = 0}^{\zeta_i}\sum_{m=1}^{M}v_{l_i m}\cdot\bigg(\bigg[\frac{d\Phibold_i(s)'}{ds} \cbold_{i}-g_{i}({\color{black} \Phibold(s)' \cbold , \Phibold(s-\tau)' \cbold}|\thetabold) \bigg]^{2}\bigg|_{s = \xi_{il_im}}\bigg)\bigg)
\bigg\}.
\end{eqnarray}
}
\item If the full conditional distribution $\pi_{r}(\thetabold |\cbold,\tau, \lambda)$ does not admit a closed form. We perform a random walk MH algorithm with a Gaussian kernel with
\begin{eqnarray}
\label{eq:thetar}
\gamma_{r}( \thetabold |\cbold,\tau, \lambda) 
\propto \exp\bigg\{ -\alpha_{r}\sum_{i=1}^{I}\bigg(\frac{\lambda}{2}\sum_{l_i = 0}^{\zeta_i}\sum_{m=1}^{M}v_{l_i m}\cdot\bigg(\bigg[\frac{d\Phibold_i(s)'}{ds} \cbold_{i}-g_{i}({\color{black} \Phibold(s)' \cbold , \Phibold(s-\tau)' \cbold}|\thetabold) \bigg]^{2}\bigg|_{s = \xi_{il_im}}\bigg)\bigg)
\bigg\}.
\end{eqnarray}

\item The full conditional posterior distribution of $\lambda$ is $\text{Gamma}(a_{\lambda} + \alpha_{r}\sum_{i=1}^{I}(L_{i}-2)/2, b_{\lambda}*)$,  where
\[ 
\frac{1}{b_{\lambda}*} = \frac{1}{b_{\lambda}} + \frac{\alpha_{r}}{2}\sum_{i=1}^{I}\sum_{l_i = 0}^{\zeta_i}\sum_{m=1}^{M}v_{l_i m}\cdot\bigg(\bigg[\frac{d\Phibold_i(s)'}{ds} \cbold_{i}-g_{i}({\color{black} \Phibold(s)' \cbold , \Phibold(s-\tau)' \cbold}|\thetabold) \bigg]^{2}\bigg|_{s = \xi_{il_im}}\bigg).
\]

\end{itemize}

\section{Three Monte Carlo methods for inference of parameters in Bayesian differential equations}

In this Section, we introduce three other Monte Carlo methods for inference of parameters in Bayesian differential equations. 
\subsection{Markov chain Monte Carlo targeting $\pi(\betabold)$}

The first approach we introduce is Markov chain Monte Carlo targeting $\pi(\betabold)$. We name it MCMC-spline for simplicity. 

We introduce the conditional posterior distributions for $\betabold$. The full conditional posterior distributions $\pi(\sigma_{i}^{2}|\cbold_{i})$, $\pi( \tau |\cbold, \thetabold, \lambda)$, $\pi( \thetabold |\cbold,\tau, \lambda)$, and $\pi( \cbold_{i} |\tau, \thetabold, \boldsymbol{\sigma}, \cbold_{-i}, \lambda)$ admit the following forms

\begin{itemize}
\item The full conditional distribution for $\sigma_{i}^{2}$ is 
\begin{eqnarray}
\label{eq:sigmar}
\sigma_{i}^{2}|\cbold_{i}\sim IG\bigg(g_{0}+\frac{{\color{black} J_i}}{2}, h_{0}+\frac{1}{2}\sum_{j = 1}^{{\color{black} J_i}}(y_{ij} - \Phibold_i(t_{ij})' \cbold_{i})^{2}\bigg),  ~~ \textcolor{black}{\text{ for } i \in \mathcal{I}_0}. 
\end{eqnarray}

\item The full conditional distribution $\pi( \tau |\cbold, \thetabold, \lambda)$ does not admit a closed form. 
\begin{eqnarray}
\label{eq:taur}
&&\gamma( \tau |\cbold, \thetabold, \lambda) 
\propto \\ &&\exp\bigg\{ -\sum_{i=1}^{I}\bigg(\frac{\lambda}{2}\sum_{l_i = 0}^{\zeta_i}\sum_{m=1}^{M}v_{l_i m}\cdot\bigg(\bigg[\frac{d\Phibold_i(s)'}{ds} \cbold_{i}-g_{i}({\color{black} \Phibold(s)' \cbold , \Phibold(s-\tau)' \cbold}|\thetabold) \bigg]^{2}\bigg|_{s = \xi_{il_im}}\bigg)\bigg)
\bigg\}. \nonumber
\end{eqnarray}

We perform a random walk MH algorithm with a Gaussian kernel to  propose $\tau$. 
\begin{enumerate}
\item $\tau^{\star}\sim N(\tau, \sigma_{\tau}^{2})$,
\item compute the acceptance probability 
\[
 p_{\mathit{MH}} = \min\bigg\{1, \frac{\gamma(\tau^{\star}|\cbold, \thetabold, \lambda)}{\gamma(\tau|\cbold, \thetabold, \lambda)}\bigg\},
\]
\item sample $u\sim U(0,1)$,  we accept $\tau = \tau^{\star}$ if $u<p_{MH}$, otherwise we set $\tau = \tau$.
\end{enumerate}

\item 
The existence of closed-form full conditional posterior distributions  $\pi(\theta_{d}|\cbold, \tau, \lambda)$ and $\pi(\cbold_{i} |\thetabold, \boldsymbol{\sigma},  \cbold_{-i}, \lambda)$ depends on $g_{i}$ $(i = 1, 2, \ldots, I)$. If all $g_{i}$ are linear functions of $\theta_{d}$ (or $\cbold_{i}$), there exists closed-form full conditional posterior distribution $\pi(\theta_{d}|\cbold, \tau, \lambda)$ (or $\pi(\cbold_{i} |\thetabold, \boldsymbol{\sigma},  \cbold_{-i}, \lambda)$), which is Gaussian distributed. Otherwise, we perform a random walk MH algorithm with a Gaussian kernel. 

\item If the full conditional distribution $\pi(\cbold_{i} |\tau, \thetabold, \boldsymbol{\sigma}, \cbold_{-i}, \lambda)$ does not admit a closed form. We perform a random walk MH algorithm with a Gaussian kernel with
{\small
\begin{eqnarray}
\label{eq:cr}
&&\gamma( \cbold_{i} |\tau, \thetabold, \boldsymbol{\sigma}, \cbold_{-i}, \lambda)\propto \nonumber\\
 &&\exp\bigg\{ -\bigg(\textcolor{black}{\sum_{i\in \mathcal{I}_0}}  \sum_{j=1}^{{\color{black} J_i}}\frac{(y_{ij} -  \Phibold_i(t_{ij})' \cbold_{i})^{2}}{2\sigma_{i}^{2}} +\nonumber\\
&&\textcolor{black}{\sum_{i=1}^{I} } \frac{\lambda}{2}\sum_{l_i = 0}^{\zeta_i}\sum_{m=1}^{M}v_{l_i m}\cdot\bigg(\bigg[\frac{d\Phibold_i(s)'}{ds} \cbold_{i}-g_{i}({\color{black} \Phibold(s)' \cbold , \Phibold(s-\tau)' \cbold}|\thetabold) \bigg]^{2}\bigg|_{s = \xi_{il_im}}\bigg)\bigg)
\bigg\}.
\end{eqnarray}
}
\item If the full conditional distribution $\pi(\thetabold |\cbold,\tau, \lambda)$ does not admit a closed form. We perform a random walk MH algorithm with a Gaussian kernel with
\begin{eqnarray}
\label{eq:thetar}
\gamma( \thetabold |\cbold,\tau, \lambda) 
\propto \exp\bigg\{ -\sum_{i=1}^{I}\bigg(\frac{\lambda}{2}\sum_{l_i = 0}^{\zeta_i}\sum_{m=1}^{M}v_{l_i m}\cdot\bigg(\bigg[\frac{d\Phibold_i(s)'}{ds} \cbold_{i}-g_{i}({\color{black} \Phibold(s)' \cbold , \Phibold(s-\tau)' \cbold}|\thetabold) \bigg]^{2}\bigg|_{s = \xi_{il_im}}\bigg)\bigg)
\bigg\}. 
\end{eqnarray}

\item The full conditional posterior distribution of $\lambda$ is $\text{Gamma}(a_{\lambda} + \sum_{i=1}^{I}(L_{i}-2)/2, b_{\lambda}*)$,  where
\[
\frac{1}{b_{\lambda}*} = \frac{1}{b_{\lambda}} + \frac{1}{2}\sum_{i=1}^{I}\sum_{l_i = 0}^{\zeta_i}\sum_{m=1}^{M}v_{l_i m}\cdot\bigg(\bigg[\frac{d\Phibold_i(s)'}{ds} \cbold_{i}-g_{i}({\color{black} \Phibold(s)' \cbold , \Phibold(s-\tau)' \cbold}|\thetabold) \bigg]^{2}\bigg|_{s = \xi_{il_im}}\bigg).
\]

\end{itemize}

\subsection{Markov chain Monte Carlo targeting $\pi(\thetabold, \tau, \xbold(0), \boldsymbol{\sigma}^{2})$}

The second method we introduce is classical Markov chain Monte Carlo algorithm {\color{black} targeting} $\pi(\thetabold, \tau, \xbold(0), \boldsymbol{\sigma}^{2})$. We name it MCMC-deSolve for simplicity. We follow the notation described in \emph{Section 2} of \emph{main manuscript}. 
The joint likelihood function of $\thetabold$, $\tau$, $\xbold(0)$, and $\boldsymbol{\sigma}^{2}$ can be written as
\begin{equation}
\label{eq:1}
L(\thetabold, \tau, \xbold(0), \boldsymbol{\sigma}^{2}) = \textcolor{black}{\prod_{i\in \mathcal{I}_0}}\prod_{j=1}^{J}(\sigma_{i}^{2})^{-1/2}\exp\bigg\{ -\frac{(y_{ij} - x_{i}(t_{ij}|\thetabold, \tau, \xbold(0)))^{2}}{2\sigma_{i}^{2}} \bigg\}.
\end{equation}
To construct a Bayesian framework, we assign appropriate prior distributions for model  parameters $\thetabold$, $\tau$, $\xbold(0)$, and $\boldsymbol{\sigma}^{2}$, denoted by  $\tilde\pi_{0}(\thetabold)$, $\tilde\pi_{0}(\tau)$, $\tilde\pi_{0}(\xbold(0))$, and $\tilde\pi_{0}(\boldsymbol{\sigma}^{2})$.
The full conditional posterior distribution of $\sigma_{i}^{2}$ is Inverse-Gamma distributed. 
The full conditional posterior distributions of $\thetabold$, $\tau$, and $\xbold(0)$ do not have analytical solutions. We conduct random walk MH {\color{black} algorithms} to sample new parameters. We use $\tau$ as an illustrative example. 
Conditional on samples at the $n$-th iteration $\thetabold^{(n)}$, $\xbold(0)^{(n)}$, and $\boldsymbol{\sigma}^{2(n)}$,
\begin{enumerate}
\item $\tau^{\star}\sim N(\tau^{(n)}, \sigma_{\tau}^{2})$.
\item Solve DEs numerically and obtain $\xbold(t_{ij}|\thetabold^{(n)}, \tau^{\star}, \xbold(0)^{(n)})$. 
\item Compute the acceptance probability 
\[
p_{MH} = \min\bigg\{1, \frac{\gamma(\thetabold^{(n)}, \tau^{\star}, \xbold(0)^{(n)}, \boldsymbol{\sigma}^{2(n)})}{\gamma(\thetabold^{(n)}, \tau^{(n)}, \xbold(0)^{(n)}, \boldsymbol{\sigma}^{2(n)})}\bigg\}.
\]
\item Sample $u\sim U(0,1)$, we accept $\tau^{(n+1)} = \tau^{\star}$ if $u<p_{MH}$, otherwise $\tau^{(n+1)} = \tau^{(n)}$. 
\end{enumerate}

\subsection{Sequential Monte Carlo targeting $\pi(\thetabold, \tau, \xbold(0), \boldsymbol{\sigma}^{2})$}

The third approach we introduce is a sequential Monte Carlo method that targets $\pi(\thetabold, \tau, \xbold(0), \boldsymbol{\sigma}^{2})$. We name it SMC-deSolve for simplicity. The SMC-deSolve framework is same as described in {\it Section 3} of main manuscript. 

We approximate the target distribution $\pi(\thetabold, \tau, \xbold(0), \boldsymbol{\sigma}^{2})$ in $R$ steps.
The subscript $r$ denotes the index of intermediate probability distributions. The last intermediate target distribution $\pi_{R}(\thetabold, \tau, \xbold(0), \boldsymbol{\sigma}^{2})$ is $\pi(\thetabold, \tau, \xbold(0), \boldsymbol{\sigma}^{2})$. We let $\gamma_{r}(\thetabold, \tau, \xbold(0), \boldsymbol{\sigma}^{2})$ denote the unnormalized version of $\pi_{r}(\thetabold, \tau, \xbold(0), \boldsymbol{\sigma}^{2})$.
At each step $r$, we use a list of $K$ samples to represent $\pi_{r}(\thetabold, \tau, \xbold(0), \boldsymbol{\sigma}^{2})$.
We use a sequence of annealed intermediate target distributions $\{\pi_{r}(\thetabold, \tau, \xbold(0), \boldsymbol{\sigma}^{2})\}_{0\leq r\leq R}$ to facilitate the exploration of posterior space, such that 
\[
\pi_{r}(\thetabold, \tau, \xbold(0), \boldsymbol{\sigma}^{2}) \propto  \gamma_{r}(\thetabold, \tau, \xbold(0), \boldsymbol{\sigma}^{2})= {p(\ybold|\thetabold, \tau, \xbold(0), \boldsymbol{\sigma}^{2})}^{\alpha_{r}} \tilde\pi_{0}(\thetabold, \tau, \xbold(0), \boldsymbol{\sigma}^{2}).
\]
We use same prior distributions $\tilde\pi_{0}(\thetabold, \tau, \xbold(0), \boldsymbol{\sigma}^{2})$ as the previous section.

We iterate between the weighting, propagation, and resampling to obtain the approximated intermediate target posterior distribution. We refer readers to {\it Section 3.1} for details of weighting and resampling steps. We use MCMC-deSolve moves described in previous section to propagate new particles.

\section{Real Data Analysis}
\label{sec:real}

\begin{figure}[ht]
\center
\includegraphics[scale=0.8]{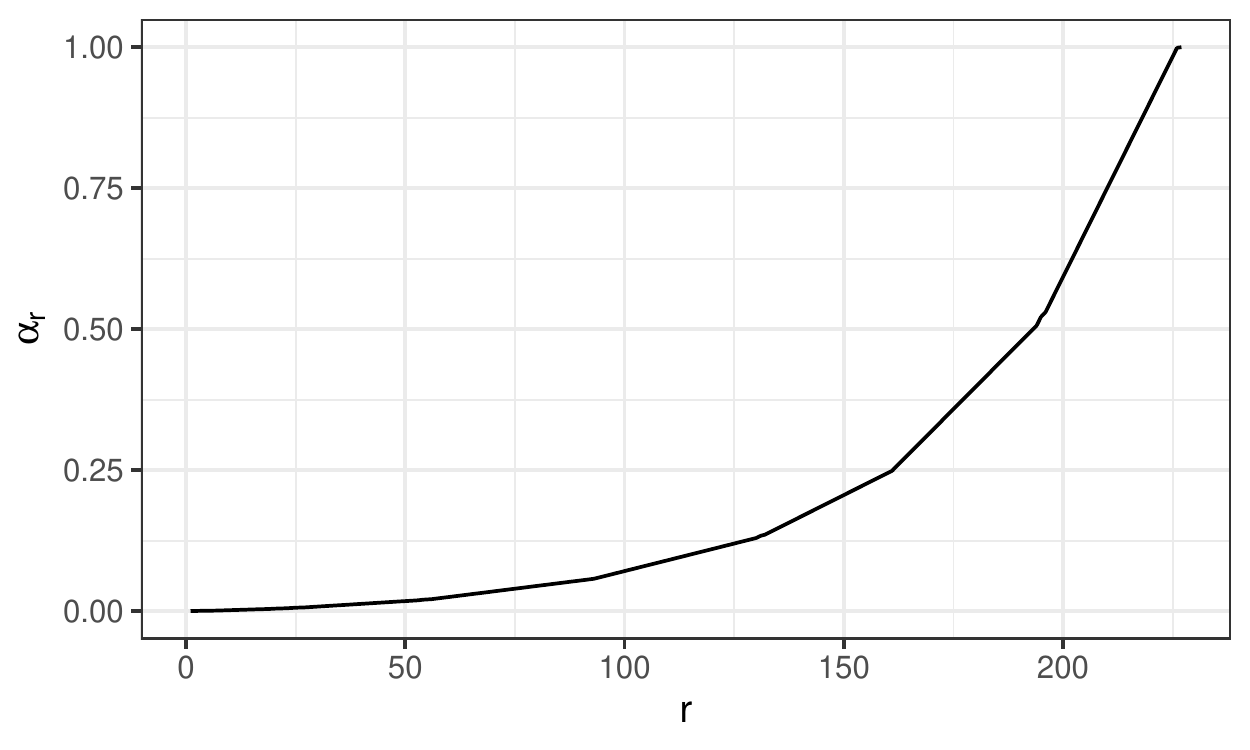}
\caption{Annealing parameter sequence $\alpha_{1:R}$. }
\label{fig:alpha}
\end{figure}

Figure \ref{fig:alpha} shows the annealing parameter sequence $\alpha_{1:R}$ under the adaptive scheme.

\section{Simulation Study}
\label{sec:simulation}
We use simulation studies to demonstrate the effectiveness of our proposed model and method. The experiments include both ODE and DDE examples. We use the R package \emph{deSolve} \citep{soetaert2010solving} to simulate differential equations. 
\subsection{A nonlinear ordinary differential equation example}
\label{sec:sim1}
\begin{figure}
\centering
\includegraphics[scale=0.8]{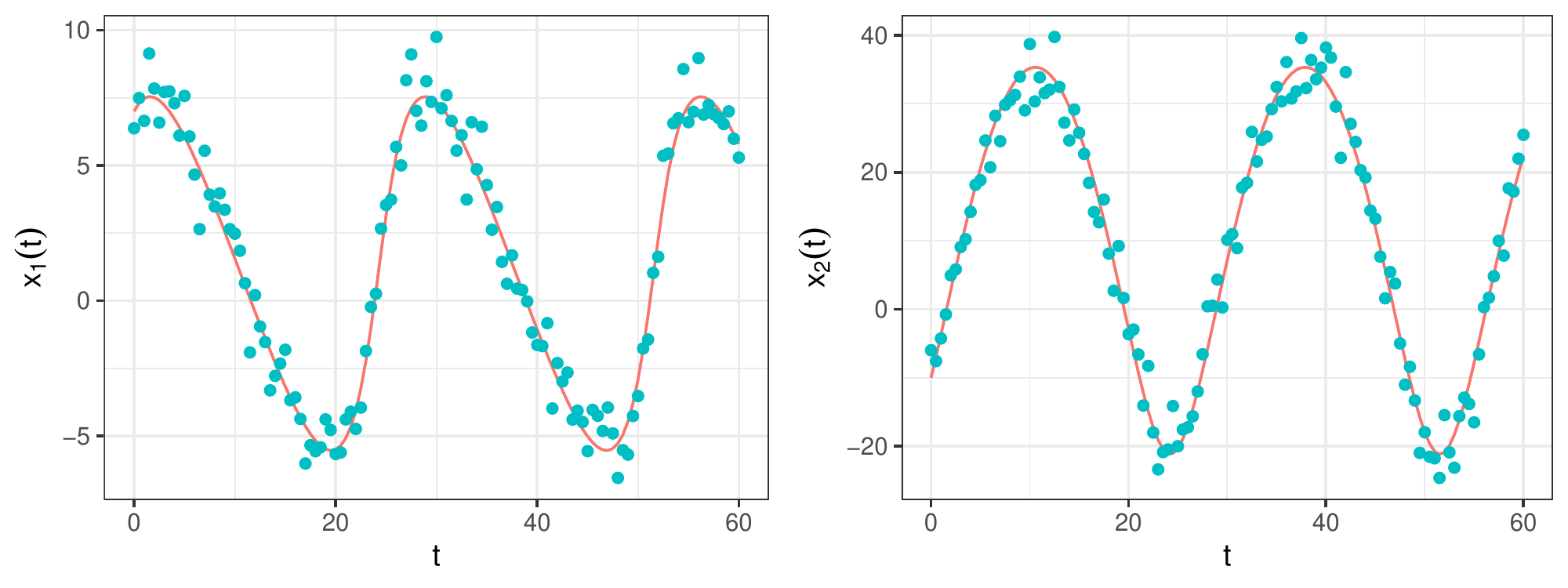}
\caption{Simulated ODE trajectories and observations. Red lines in the figure refer to simulated ODE trajectories and blue points refer to simulated observations. } 
\label{fig:ode2}
\end{figure}

In this section, we use a nonlinear ODE example to illustrate the numerical behaviour of the SMC algorithm.  We generate ODE trajectories according to the following ODE system,
\begin{eqnarray}
\label{eq: ode1}
\frac{dx_{1}(t)}{dt} &=& \frac{72}{36+x_{2}(t)} - \theta_{1},\nonumber\\
\frac{dx_{2}(t)}{dt} &=& \theta_{2}x_{1}(t) - 1,
\end{eqnarray}
where $\theta_{1} = 2$ and $\theta_{2} = 1$, and initial conditions $x_{1}(0) = 7$ and $x_{2}(0) = -10$. The observations $\ybold_{i}$ are simulated from a normal distribution with mean  $x_{i}(t|\thetabold)$ and variance $\sigma_{i}^{2}$, where $\sigma_{1} = 1$ and $\sigma_{2} = 3$. We generate $121$ observations for each ODE function, equally spaced within $[0, 60]$ (see Figure \ref{fig:ode2}).

We use cubic B-spline functions (see Figure \ref{fig:basisfunction}) to represent ODE trajectories. We put equally spaced knots on each of eight observations. The total number of knots is $16$, including $14$ interior knots. The total number of cubic B-spline functions is ${\color{black} L_1 = L_2} = 18$. 
We select weak  prior/reference distributions of $\betabold$ for the SMC algorithm,
 \begin{eqnarray*}
\theta_{1} &\sim& \text{N}(5,5^2), ~~~\theta_{2} ~\sim~ \text{N}(5,5^2), \\
 \cbold_{1} &\sim& \text{MVN}(\hat{\cbold}_{1},100^2\Ibold_{{\color{black} L_1}}),
 ~~~\cbold_{2} ~\sim~ \text{MVN}(\hat{\cbold}_{2},100^2\Ibold_{{\color{black} L_2}}),\\
 \sigma_1^2 &\sim& \text{IG}(1, 1), ~~~\sigma_2^2 ~\sim~ \text{IG}(1, 1), ~~~\lambda ~\sim~ \text{Gamma}(1, 1).
 \end{eqnarray*}

\subsubsection{One bimodal example}
We first alter Equation (\ref{eq: ode1}) to produce a symmetric, bimodal
posterior for $\theta_{1}$, 
\begin{eqnarray}
\label{eq: ode2}
\frac{dx_{1}(t)}{dt} &=& \frac{72}{36+x_{2}(t)} - |\theta_{1}|,\nonumber\\
\frac{dx_{2}(t)}{dt} &=& \theta_{2}x_{1}(t) - 1.
\end{eqnarray}

\begin{figure}
\centering
\includegraphics[scale=0.7]{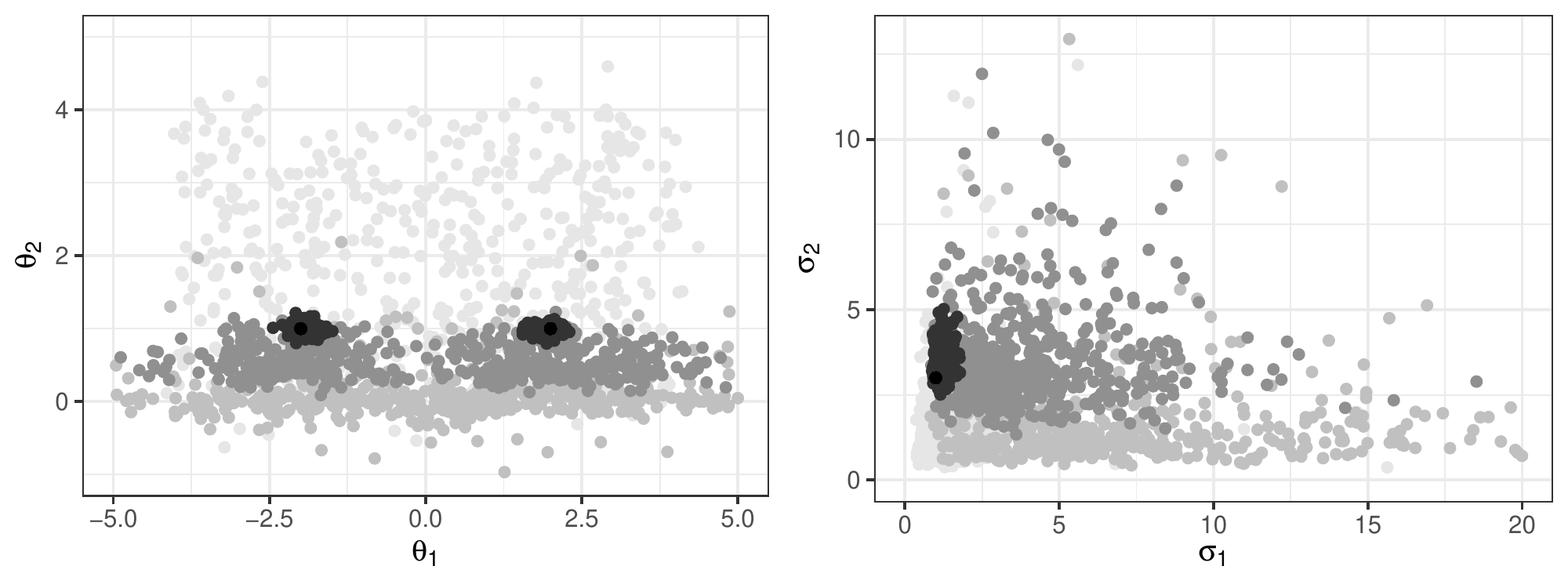}
\includegraphics[scale=0.7]{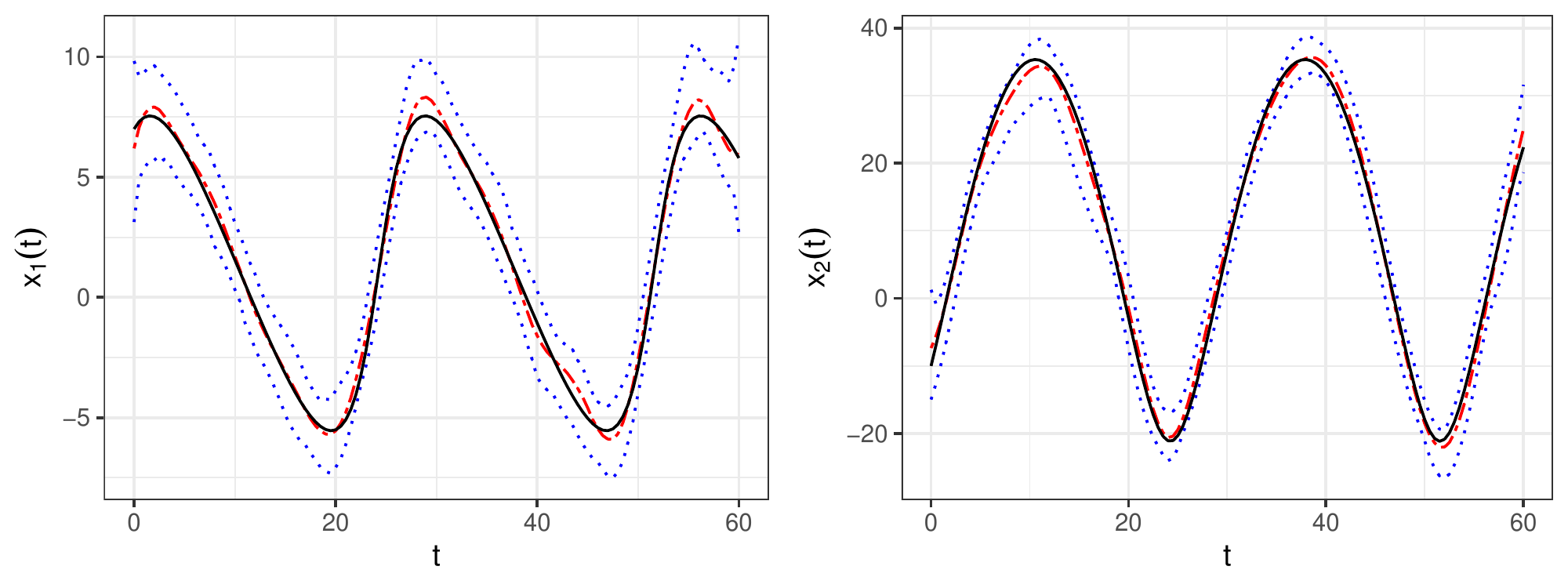}
\caption{Intermediate posterior distributions for $\thetabold$ and $\boldsymbol{\sigma}$ (two panels in the first row, {\color{black} points with color from light grey to dark grey}  are samples for  $r = 1, R/6, R/2, R$, the black dots indicate true values for generating ODE), and estimated ODE trajectories and the 95\% confidence bands (two panels in the second row). }
\label{fig:para}
\end{figure}

In our adaptive SMC, we set the rCESS threshold $\phi = 0.9$ and resampling threshold $\varsigma = 0.5$. The total number of particles we use is $K = 500$. 
Under this setting, the number of annealing parameters is $752$. We show the approximated intermediate posterior distributions for $\thetabold$ and $\boldsymbol{\sigma}$ {\color{black} for $r = 1, R/6, R/2, R$ with color from light grey to dark grey (see two panels in the first row of Figure \ref{fig:para}).}
With the increment of annealing parameters, the particles gradually move to the posterior distribution. We create two main modes for $\theta_{1}$ in Equation (\ref{eq: ode2}).  The SMC algorithm is able to find these two global modes of $\theta_{1}$ while avoiding being stuck in local modes. 
We {\color{black} report} the estimated ODE trajectories and the $95\%$ {\color{black} pointwise} credible bands in the two bottom panels  of {Figure \ref{fig:para}}. The estimated mean ODE trajectories are very close to the true ODE trajectories. The $95\%$ {\color{black} pointwise} credible bands covers the true ODE trajectories.

\subsubsection{Comparison of SMCs using different $\phi$ and $K$ }
\label{sec:Comparison of using different rCESS and K}
We conduct experiments to investigate the performance of SMC algorithm with different thresholds of $\phi$ and $K$ using the ODE system showed in Equation (\ref{eq: ode1}). We simulate $20$ datasets according to  {\it Section} \ref{sec:sim1}. We compare the performance of SMC algorithm in terms of estimated $\thetabold$, $\boldsymbol{\sigma}$, and estimated ODE. With estimated basis coefficients $\hat{\cbold}_{i}$, we are able to compute the estimated $i$-th ODE trajectory $\hat{x}_{i}(t) = \Phibold_i(t)' \hat{\cbold}_{i}$. We define the distance between the estimated $i$-th ODE trajectory $\hat{x}_{i}(t)$ and the true ODE $x_{i}(t)$ that we use to simulate data as 
\begin{equation}
\label{eq:rmse}
\text{RMSE}(x_{i}(t)) = \bigg[\frac{1}{{\color{black} J_i}}\sum_{j=1}^{{\color{black} J_i}} (\Phibold_i(t_{ij})' \hat{\cbold}_{i} - x_{i}(t_{ij}))^{2} \bigg]^{1/2}.
\end{equation}

\begin{figure}[ht]
\center
\includegraphics[scale=0.7]{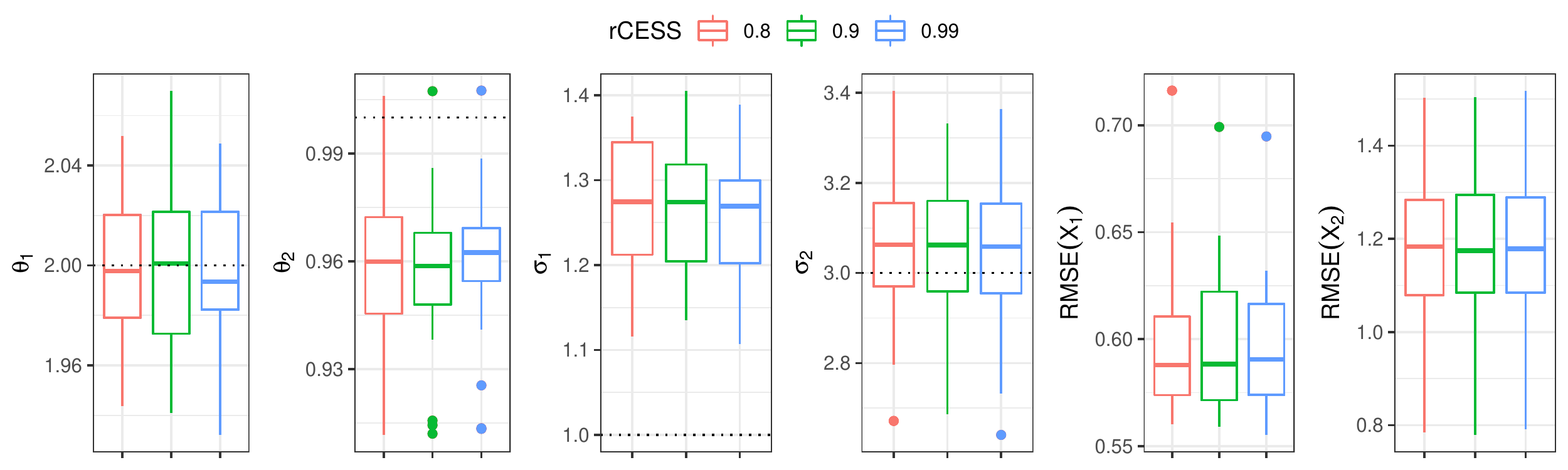}
\caption{ODE parameter estimates with different rCESS in SMC. }
\label{fig:R}
\end{figure}

We select three different levels of rCESS threshold $\phi$ ($\phi = 0.8, 0.9, 0.99$). 
We put equally spaced interior knots on each of the $12$ observations. The total number of basis function is $13$.  
The number of particles we use is $K = 500$.  For each level of $\phi$, we run the SMC algorithm $20$ times (once for each dataset).  Figure \ref{fig:R} displays boxplots of the posterior samples of 
$\thetabold$, $\boldsymbol{\sigma}$, and $\text{RMSE}(x_{i}(t))$ from 
SMC with different rCESS thresholds. It indicates that the parameter estimates get closer to the true values, and RMSE of the ODE trajectories gets smaller, when we increase the rCESS threshold. A higher value of rCESS threshold is equivalent to more intermediate target distributions.

\begin{figure}[ht]
\center
\includegraphics[scale=0.7]{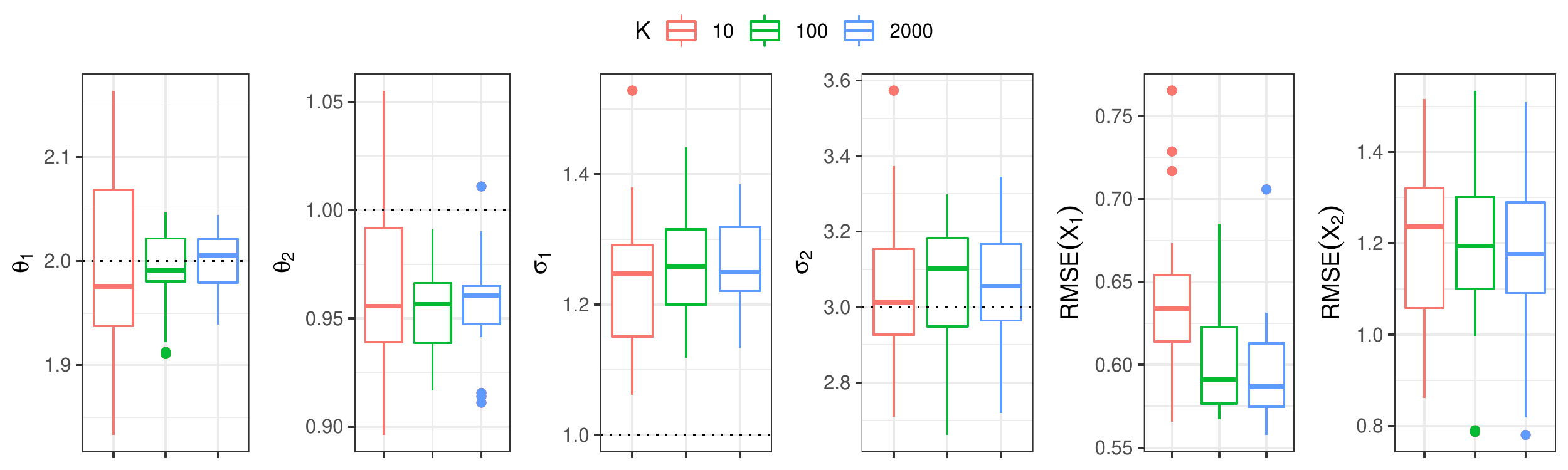}
\caption{Boxplots of ODE parameter estimates with different values of $K$ in SMC. }
\label{fig:K}
\end{figure}

We select three different levels of $K$ ($K = 10, 100, 2000$). 
We also put equally spaced interior knots and the total number of basis function is $13$.  
We set rCESS threshold $\phi = 0.9$. For each level of $K$,  we run the SMC algorithm $20$ times  (once for each dataset).  Figure \ref{fig:K} displays boxplots for $\thetabold$, $\boldsymbol{\sigma}$, and $\text{RMSE}(x_{i}(t))$ from SMC with different levels of $K$. It indicates that the proposed SMC method performs better when we use a large number of particles.  The consistency of the SMC algorithm holds when the number of particles goes to infinity \citep{chopin2004central, wang2018annealed, del2006sequential}. However, we cannot use an arbitrarily large value of $K$ as the computational cost of the SMC algorithm is a linear function of $K$. We recommend increasing $\phi$ in SMC (using a larger number of intermediate distributions $R$), as increasing $R$ does not increase the memory burden. 

 In addition, we conducted an experiment to investigate the
relative importance of $\phi$ and $K$ in improving the quality of estimation by
SMC. We selected four levels of $K$, $\phi$ combinations: $(K, \phi) = (10, 0.99), (100, 0.8), (1000, 0.08), (6000, 0.001)$. For each value of $K$, we select a corresponding value
for $\phi$ such that the total computation cost $K\cdot R$ is close to $80,000$. 
We simulate $20$ datasets according
to {\it Section \ref{sec:sim1}}.
 Figure 7
displays boxplots of ODE parameter estimates with different combination of $K$, $\phi$ values in SMC. This results indicate that for a given
amount of computation, a relatively small $K$ and a large $\phi$ is optimal. However, a too small value of $K$ is not recommended, as an extremely small $K$ may lead to a large Monte Carlo variance. 

\begin{figure}[ht]
\center
\includegraphics[scale=0.7]{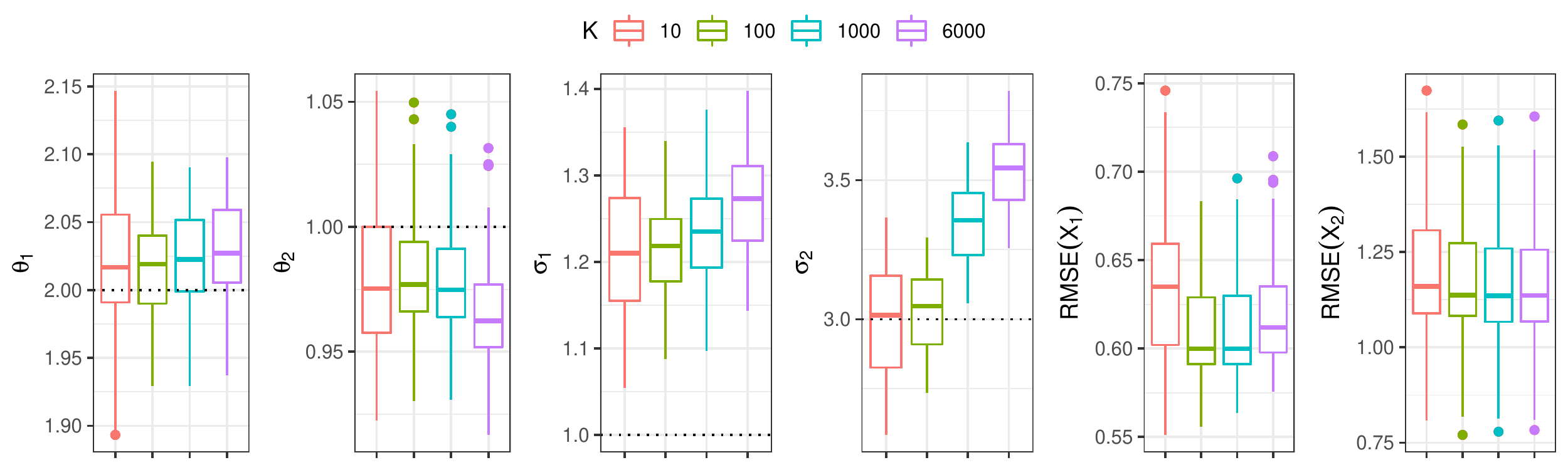}
\caption{Boxplots for ODE parameter estimates with different values combination of $K$, $\phi$ values in SMC. }
\label{fig:trade-off}
\end{figure}

\subsubsection{Number of basis functions and selection of $\lambda$}
\label{subsection:Number of basis functions and selection of lambda}
\begin{figure}[ht]
\center
\includegraphics[scale=0.7]{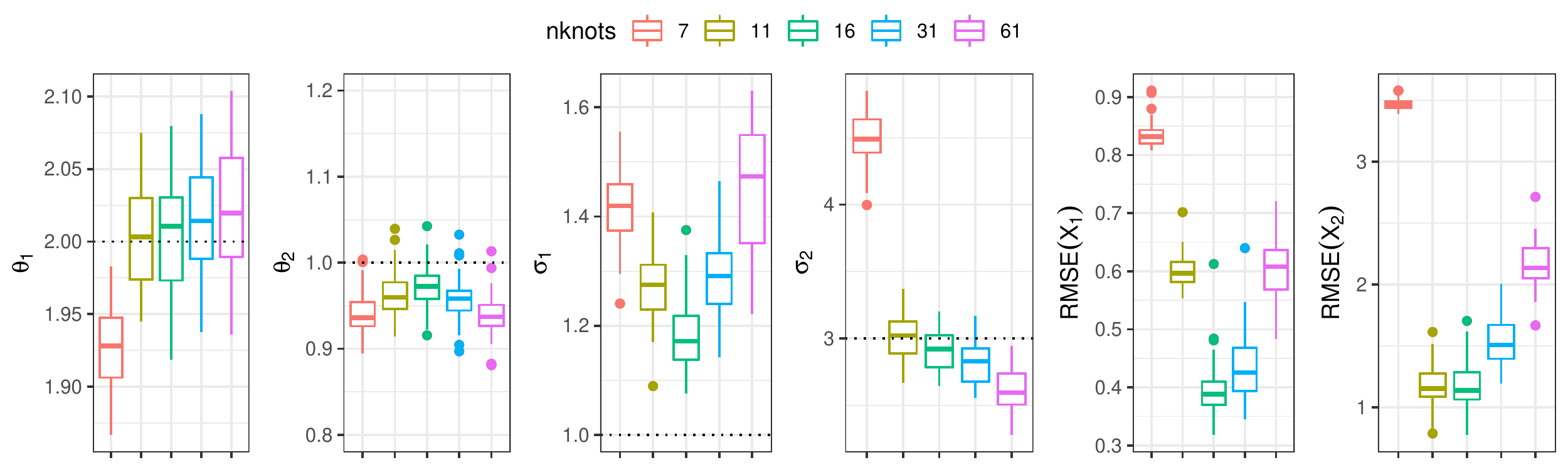}
\caption{ODE parameter estimates obtained from SMC with different numbers of knots. }
\label{fig:nknots}
\end{figure}

In this section, we first conduct experiments to investigate the performance of SMC algorithm with different numbers of basis functions in terms of estimating $\thetabold$, $\boldsymbol{\sigma}$ and the RMSE of the estimated ODE trajectories. The knots for basis function are equally spaced. We choose five different numbers of basis function, $\text{nbasis} = 7, 11, 16, 31, 61$. We set  $K = 1000$ and  the rCESS threshold $\phi = 0.95$ for SMC. For each level of number of basis function, we run SMC once for each of the 40 datasets simulated according
to {\it Section \ref{sec:sim1}}. Figure \ref{fig:nknots} displays the ODE parameter estimates with different numbers of basis functions. 
The parameter estimates $\thetabold$ and $\boldsymbol{\sigma}$ get closer to true values, and the RMSE values of estimated ODE trajectories decrease  if we increase the number of basis functions from $7$ to $16$.  However, the parameter estimates and RMSE  of the estimated ODE trajectories become worse if we use  a large number of basis functions. This experiment indicates a sufficient number of basis functions is important in ODE trajectory estimation. However, we do not recommend using a overly large number of basis functions as it will cause over fitting and induce a heavy computational cost.

\begin{figure}[ht]
\center
\includegraphics[scale=0.7]{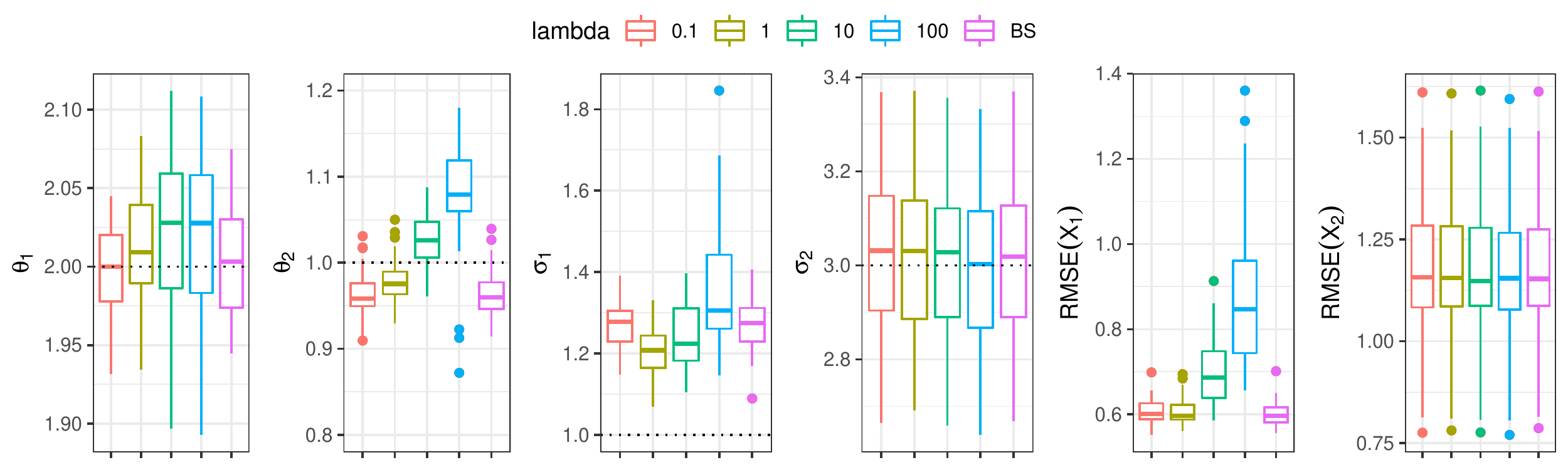}
\caption{ODE parameter estimates with different choices of $\lambda$. }
\label{fig:lambda}
\end{figure}

\begin{figure}[ht]
\center
\includegraphics[scale=0.7]{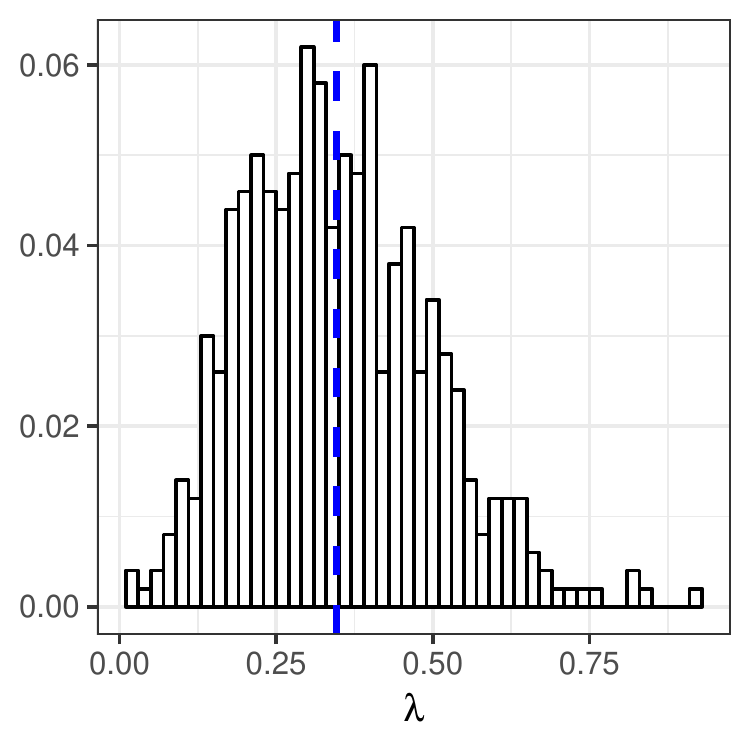}
\caption{One example of posterior samples of $\lambda$. }
\label{fig:lambda_post}
\end{figure}

The second experiment we conduct is a comparison between the performance of SMC algorithm with different choice of $\lambda$ ($\lambda = 0.1, 1, 10, 100$) and the Bayesian scheme (BS). We put equally spaced knots and set the number of basis functions to $16$. We set $K = 1000$ and the rCESS threshold $\phi = 0.95$ for  SMC algorithm. For each  choice of $\lambda$, we run the SMC algorithm with one replicate for the 40 datasets simulated in according
to  {\it Section \ref{sec:sim1}}. Figure \ref{fig:lambda} displays the ODE parameter estimates with different choices of $\lambda$. The Bayesian scheme  performs satisfactorily in terms of parameter estimates and RMSE of the ODE trajectories. Figure \ref{fig:lambda_post} displays the posterior samples of $\lambda$ for one SMC  replicate. 

\subsection{A nonlinear delay differential equation example}
In this section, we investigate a nonlinear delay differential equation model proposed by \cite{monk2003oscillatory} to model the feedback inhibition of gene expression. The nonlinear DDE is described as follows:
\begin{eqnarray}
\label{eq: ndde}
\frac{dx_{1}(t)}{dt} & = & \frac{1}{1+(x_{2}(t-\tau)/p_{0})^{n}}-\mu_{m}x_{1}(t), \nonumber\\
\frac{dx_{2}(t)}{dt} & = & x_{1}(t)-\mu_{p}x_{2}(t).
\end{eqnarray}
In Equation (\ref{eq: ndde}), $x_{1}(t)$ denotes the expression of \emph{mRNA} at time $t$, and $x_{2}(t)$ denotes the expression of a \emph{protein} at time $t$. There is a delayed repression of \emph{mRNA} production by the \emph{protein}. The DDE system depends on the \emph{transcriptional delay} $\tau$, and degradation rates $\mu_{m}$ and $\mu_{p}$, the expression threshold $p_{0}$ and the Hill coefficient $n$. As noted in \cite{monk2003oscillatory}, there is significant nonlinearity in the DDE system when the Hill coefficient $n > 4$.

\begin{figure}
\centering
\includegraphics[scale=0.8]{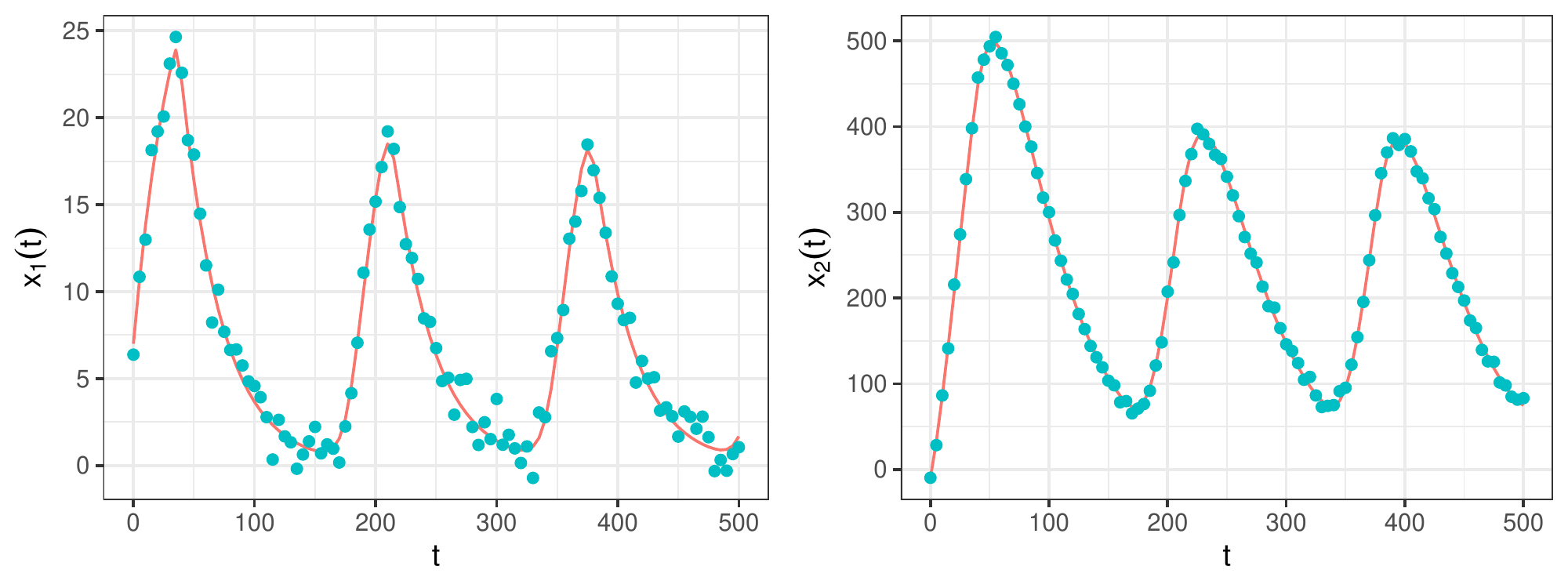}
\caption{Simulated DDE trajectories and observations. Red lines in Figure refer to simulated DDE trajectory and blue points refer to simulated observations.}
\label{fig:2}
\end{figure}

We simulate a delay differential equation system with $\tau = 25$, $p_{0} = 100$, $\mu_{m} = 0.03$, $\mu_{p} = 0.03$,  and $n$ is set to $8$.  The observations $y_{i}(t)$ are simulated from a normal distribution with mean  $x_{i}(t|\thetabold)$ and variance $\sigma_{i}^{2}$, where $\sigma_{1} = 1$ and $\sigma_{2} = 5$. We generate $101$ observations for each DDE function, equally spaced in $[0, 500]$. Figure \ref{fig:2} represents the simulated DDE system, which exhibits oscillations in \emph{mRNA} and \emph{protein}  expression. 

%
%

We allocate equally spaced knots within $[0, 500]$. The total number of cubic B-spline basis function is $L = 28$.  We select the weak prior/reference distributions of $\betabold$ for the SMC algorithm,
 \begin{eqnarray*}
\theta_{1} &\sim& \text{N}(0,5^2), 
~~~\theta_{2} ~\sim~ \text{N}(0,5^2), ~~~\tau \sim \text{Unif}(0,50), \\
 \cbold_{1} &\sim& \text{MVN}(\hat{\cbold}_{1},100^2\Ibold_{{\color{black} L_1}}),
~~~ \cbold_{2} ~\sim~ \text{MVN}(\hat{\cbold}_{2},100^2\Ibold_{{\color{black} L_2}}),\\
 \sigma_1^2 &\sim& \text{IG}(1, 1),
 ~~~ \sigma_2^2 ~\sim~ \text{IG}(1, 1),  ~~~ \lambda ~\sim~ \text{Gamma}(1, 1).
 \end{eqnarray*}
In our adaptive SMC, we set  $\phi = 0.9$ and resampling threshold $\varsigma = 0.5$. The total number of particles we use is $K = 300$. 
Under this setting, the number of annealing parameters is $R = 850$. 
We show the parameter estimates and  the corresponding $95\%$ credible interval (CI) in Table \ref{tab:2}. The mean of parameters are fairly close to the true values, and the 95\% credible intervals cover the true values. The estimated posterior mean of $\lambda$ is $0.225$.

%

\begin{table}[ht]
\centering
\caption{Parameter estimates and the corresponding $95\%$ credible interval (CI) for nonlinear DDE models.}
\label{tab:2}
\begin{tabular}{rrrr}
  \hline
  &True& Mean & 95\% CI \\ 
  \hline
$\mu_{m}$ & 0.03& 0.028& (0.008, 0.051) \\ 
 $\mu_{p}$& 0.03& 0.030& (0.029, 0.032) \\ 
 $p_{0}$& 100& 94.97 & (68.49, 120.06) \\ 
  $\tau$& 25& 24.74 & (12.38, 34.74) \\ 
 $\sigma_{1}$ & 1&1.02& (0.87, 1.17) \\ 
 $\sigma_{2}$ & 5&4.58& (4.04, 5.25) \\ 
  $x_{1}(0)$ & 7&6.87& (4.39, 9.14) \\ 
 $x_{2}(0)$ & -10&-10.75& (-16.66, -5.06) \\ 

   \hline
\end{tabular}
\end{table}

\begin{figure}
\centering
\includegraphics[scale=0.7]{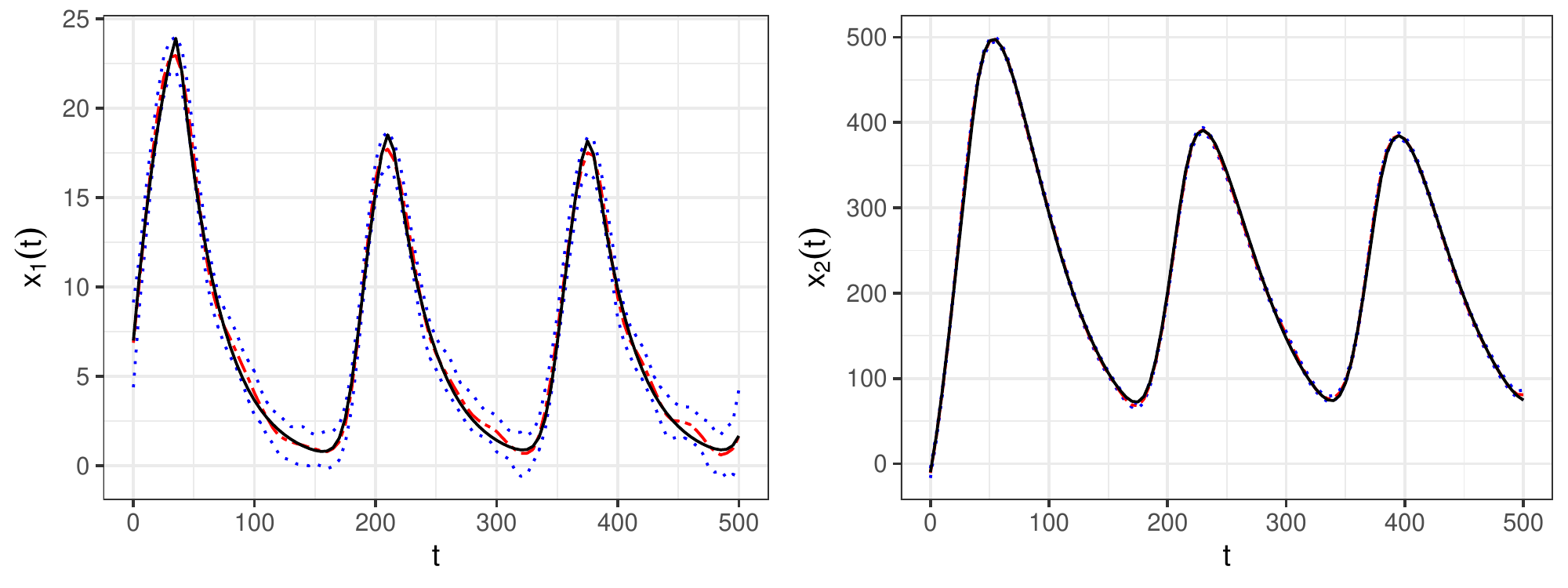}
\caption{Estimated DDE trajectories and the 95\% confidence bands using $L=28$ basis functions.}
\label{fig:5}
\end{figure}

We reported the estimated DDE trajectories and the $95\%$ {\color{black} pointwise credible intervals}  in Figure \ref{fig:5}. The estimated mean DDE trajectories are generally very close to the true DDE trajectories.  The $95\%$ {\color{black} pointwise credible intervals} cover the true DDE trajectories. 


